\documentclass[aps,prb,twocolumn,showpacs]{revtex4-1}
\usepackage{amssymb}
\usepackage{graphicx}
\usepackage{amsmath}

\usepackage{times}
\usepackage{color}
\usepackage{subfigure}
\usepackage{setspace}
\usepackage{epstopdf}
\usepackage{bm}

\begin{document}

\newcommand {\beq} {\begin{equation}}
\newcommand {\eeq} {\end{equation}}
\newcommand {\bqa} {\begin{eqnarray}}
\newcommand {\eqa} {\end{eqnarray}}
\newcommand {\ba} {\ensuremath{b^\dagger}}
\newcommand {\Ma} {\ensuremath{M^\dagger}}
\newcommand {\psia} {\ensuremath{\psi^\dagger}}
\newcommand {\psita} {\ensuremath{\tilde{\psi}^\dagger}}
\newcommand{\lp} {\ensuremath{{\lambda '}}}
\newcommand{\A} {\ensuremath{{\bf A}}}
\newcommand{\Q} {\ensuremath{{\bf Q}}}
\newcommand{\kk} {\ensuremath{{\bf k}}}
\newcommand{\qq} {\ensuremath{{\bf q}}}
\newcommand{\kp} {\ensuremath{{\bf k'}}}
\newcommand{\rr} {\ensuremath{{\bf r}}}
\newcommand{\rp} {\ensuremath{{\bf r'}}}
\newcommand {\ep} {\ensuremath{\epsilon}}
\newcommand{\nbr} {\ensuremath{\langle ij \rangle}}
\newcommand {\no} {\nonumber}
\newcommand{\up} {\ensuremath{\uparrow}}
\newcommand{\dn} {\ensuremath{\downarrow}}
\newcommand{\rcol} {\textcolor{red}}

\begin{abstract}
We propose an experimental setup using  ultracold atoms to implement a bilayer
honeycomb lattice with Bernal stacking. In
presence of a potential bias between the layers and at low densities,
Fermions placed in this lattice form an annular Fermi
sea. The presence of two Fermi surfaces leads to interesting
patterns in Friedel oscillations and RKKY interactions in presence of
impurities. Furthermore, a repulsive fermion-fermion interaction leads to a Stoner 
instability towards an incommensurate spin-density-wave order with a
wave vector equal to the thickness of the Fermi sea. The instability occurs
at a critical interaction strength which goes down with the density of the
fermions. We find that the instability survives interaction
renormalization due to vertex corrections and discuss how this can be
seen in experiments. We also track the renormalization group flows of
the different couplings between the fermionic degrees of freedom, and
find that there are no perturbative instabilities, and that Stoner
instability is the strongest instability which occurs at a critical threshold
value of the interaction. The critical interaction goes to zero as the chemical potential
is tuned towards the band bottom.
 \end{abstract}
\title{Bilayer honeycomb lattice with ultracold atoms: Multiple Fermi
  surfaces and incommensurate spin density wave instability}
\author{Santanu Dey$^1,^2$ and Rajdeep Sensarma$^3$}
 \affiliation{$^1$ Department of Astronomy and Astrophysics, Tata Institute of Fundamental
 Research, Mumbai 400005, India\\
$^2$ Institut f\"{u}r Theoretische Physik, Technische Universit\"{a}t Dresden, 01062 Dresden, Germany\\
$^3$ Department of Theoretical Physics, Tata Institute of Fundamental
 Research, Mumbai 400005, India.}

\date{\today}

\maketitle
\section{Introduction}
The presence of multiple Fermi surfaces is a common and recurring
theme in electronic systems, occuring in systems as varied as simple
metals like $\text{Ti}$ or $\text{Cr}$, to doped topological insulators like $\text{Nb}$
doped $\text{Bi}_2\text{Se}_3$~\cite{dopedtopo} to heavy
fermion compounds like $\text{UPt}_3$~\cite{UPT3} to recently discovered Iron
based high temperature superconductors~\cite{FEAS,pnictide}. In some cases, they simply add an
additional quantum number to the low energy theory and change the low
energy behavior of the system only in quantitative aspects.  In more complicated systems like iron-
based superconductors ~\cite{pnictide,FEAS}, the presence of almost nested Fermi surfaces are believed
to play a more complex role. The interaction between electrons on the
different Fermi surfaces leads to strong spin fluctuations at the
nesting wave vector, which in turn drives a superconducting
instability in the system. It is, therefore, interesting to study the
complex interplay of multiple Fermi surfaces and interactions in a
system where one can change both the interaction scales and the shape
or size of the Fermi surfaces in a controllable way.

Graphene and its few layer counterparts have emerged as two-dimensional (2D) 
systems where the shape of the dispersion as well as the
size of the Fermi surface can be tuned controllably by using gate
voltages in various configurations. Bilayer graphene, which consists of two layers of carbon atoms arranged in
a honeycomb lattice in a Bernal AB stacking arrangement is an
interesting platform to study the effects of electron-electron
interactions in 2D chiral systems~\cite{blg1,blg2}. The tunability of
the carrier
density by gating the system, together with the low energy bands with
quadratic dispersion,  provide access to strongly interacting
regimes at low carrier density. The strong interaction can lead to different symmetry broken
ground states~\cite{vafek,macdonald} and non-Fermi liquid
behavior~\cite{nfl,sensarma1} at the charge neutrality point in this
system. However, in real materials, some of these effects may be hard to
observe due to the presence of strong Coulomb and short-range
impurities~\cite{disorder} which lead to the formation of puddles of
positively and negatively charged regions even in a sample that is
charge neutral on average.

In biased bilayer graphene, an effective
electric field between the layers breaks the layer symmetry and the potential
difference between the layers gaps out the low lying bands, with a
tunable band gap which can be controlled by the gate
voltages~\cite{bbilayer_expt}. More recently, it has been proposed
that domain walls between $A-B$ and $B-A$ type biased bilayer graphene
can sustain a pair of topological edge states with different valley
quantum numbers~\cite{bilayer_topo}. In a homogeneous biased bilayer
graphene, the low
energy dispersion of the bands has a sombrero (Mexican hat) like
dispersion, with the band bottom forming a circle in the Brillouin
zone. If the chemical potential is tuned to lie in the well of the
Mexican hat, the Fermi sea takes the shape of an annulus. 
 The presence of 2 Fermi surfaces, a diverging density of
states at the band bottom, and strong electronic correlations are
expected to lead to a wide range of
interesting phenomena in this system. However, in material bilayer graphene, the
depth of the Mexican hat well is comparable to the energy scale of disorder
in current samples ~\cite{bbilayer_expt,disorder}, and thus the
phenomena due to the presence of 2 Fermi surfaces will be smeared out in these systems. 

Ultracold atoms have emerged in recent years as a new platform to
study interacting quantum many body systems~\cite{Bloch_review,Esslinger_review}. The
precise knowledge of tunable Hamiltonian parameters and absence of
disorder in these systems has made them ideal for controlled access to various
phases of matter like Mott insulators, superfluids etc. and the phase
transitions between them~\cite{Greiner}. Thus, an implementation of
the biased Bernal-stacked bilayer honeycomb lattice with ultracold
atomic systems will provide the opportunity to study the interplay of
multiple Fermi surfaces and interaction, since each can be
individually controlled in this system. These systems are ideal for probing the
physics described above as one can use the tunability of parameters to make the sombrero well deeper, and the cleanliness of the
system precludes disorder washing out the phenomenology.

In this paper, we propose an
experimental setup of ultracold atoms to implement a Bernal-stacked bilayer honeycomb lattice with a
potential bias between the layers and show that there is a wide range
of experimentally achievable parameters which allow the observation
of the effects of the annular Fermi sea with two Fermi surfaces.  We
show that, at low densities of ultracold
fermions, the presence of two Fermi surfaces
leads to interesting
patterns of density oscillations (Friedel oscillations) in the presence of
impurities. Thus the presence of the annular sea and its consequences
can be captured even for non-interacting fermions. In ultracold atomic
systems the fermions interact through a short-range interaction
characterized by an s-wave scattering length in the continuum. This
leads to a Hubbard model on the lattice, where the sign of the
interaction can be tuned through a Feshbach resonance. We note that since we are interested in a lattice model
close to half-filling, the usual concerns of rapid atom loss in
continuum repulsive systems do not play a role in our implementation,
and the interacting system is stable to major atom loss. Repulsive interactions
between fermions on the biased bilayer honeycomb lattice lead to an instability towards an incommensurate
spin-density-wave (SDW) state with a wave vector which is equal to the thickness of the
Fermi sea. We find that this Stoner instability
survives the manybody renormalization of the interaction with small
changes in the critical coupling. A renormalization group analysis of
the problem at low energies shows that there are no perturbative
instabilities in this system which can overshadow the Stoner
instability (which is a threshold phenomenon, as it requires a finite
interaction strength); and the SDW is the strongest instability
within a finite threshold mean-field analysis among competing orders.
We discuss several possible experimental
signatures of the new order. We note that a prediction of Stoner
instability at $q=0$ exists for biased bilayer
graphene~\cite{bbilayer_ferro}, but this is different from our
prediction of an incommensurate spin-density wave in cold atoms with
short range interactions.

The rest of the paper is organized as follows: (i) In Sec.
~\ref{sec:cold}, we first describe our proposal for experimental
implementation of the biased bilayer honeycomb lattice with Bernal
stacking using ultracold atoms. We then describe the band dispersion
in such a system and relate these to the optical lattice parameters
using a band structure calculation. We show that there is a wide range
of experimental parameters where the effects of the annular Fermi sea
can be seen and comment on the optimal parameters for the
experiments. (ii) In Sec. ~\ref{sec:pol}, we consider the static
polarizability of a system of non-interacting fermions in this lattice, which governs
its response to potential perturbations. We calculate this using (a) the
detailed band dispersion and the band wavefunctions, and (b) a one-band
approximate dispersion, which provides analytic insight into the
problem. We find that there are three singularities of this function,
which lead to the presence of three wave vectors in the Friedel
oscillations in these systems. (iii) In Sec. ~\ref{sec:Stoner}, we
consider the possibility of an incommensurate spin-density wave
instability in a system of repulsively interacting fermions in this
lattice. We first consider the simple Stoner picture, where a
spin-density wave instability occurs at a critical coupling
monotonically decreasing with density. We further show that the Stoner
instability is almost unchanged if we go beyond Stoner approximation
and include vertex corrections which incorporate reduction of the
interaction due to many body effects. (iv) In Sec. ~\ref{sec:RG},
we consider the possibility of competing instabilities within a
renormalization group framework. We show that in this system, there is
no perturbative instability which overtakes the spin density wave
instability mentioned above. Furthermore, within mean field analysis, the
Stoner instability is the strongest among different orders quadratic
in the Fermi fields. This strengthens the case for the SDW to be the
leading instability in the system. We finally conclude in Sec.
~\ref{sec:conclusion} with a discussion and outlook on this problem.

\section{\label{sec:cold}Bilayer Honeycomb lattice with cold atoms} 
Ultracold atoms in optical lattices, where a periodic potential is
created by standing waves formed by lasers, have been used to study
strongly interacting manybody systems in a controllabel way~\cite{Bloch_review}.
In this section, we will first present a scheme to implement biased
Bernal stacked bilayer honeycomb lattice with ultracold atomic
systems in optical lattices. We will then look at the dependence of the important tight
binding parameters on the optical lattice parameters using a band
structure calculation. From the band dispersion, we will extract the
range of tight binding parameters and chemical potentials which are
optimal for observing the effects of the two Fermi surfaces and the
consequent Stoner instability. Finally, we will make the connection to
the optical lattice parameters through our band structure calculation
and show that there is a wide range of experimentally accessible
parameters where the effects of the presence of two Fermi surfaces can
be seen.

\subsection{The Optical Lattice}
A honeycomb lattice can be formed by interfering three 
coplanar laser beams propagating at an angle $\pm \pi/3$
with respect to each other ~\cite{Demler_Kitaev_2003}.
Variants of this scheme has been experimentally
implemented~\cite{Taruell_Honeycomb} to observe the low energy Dirac
quasiparticles. While there are proposals to
create A-A bilayers~\cite{bilayer_AB_prop}, 
\begin{figure}
\includegraphics[width =0.45\linewidth]{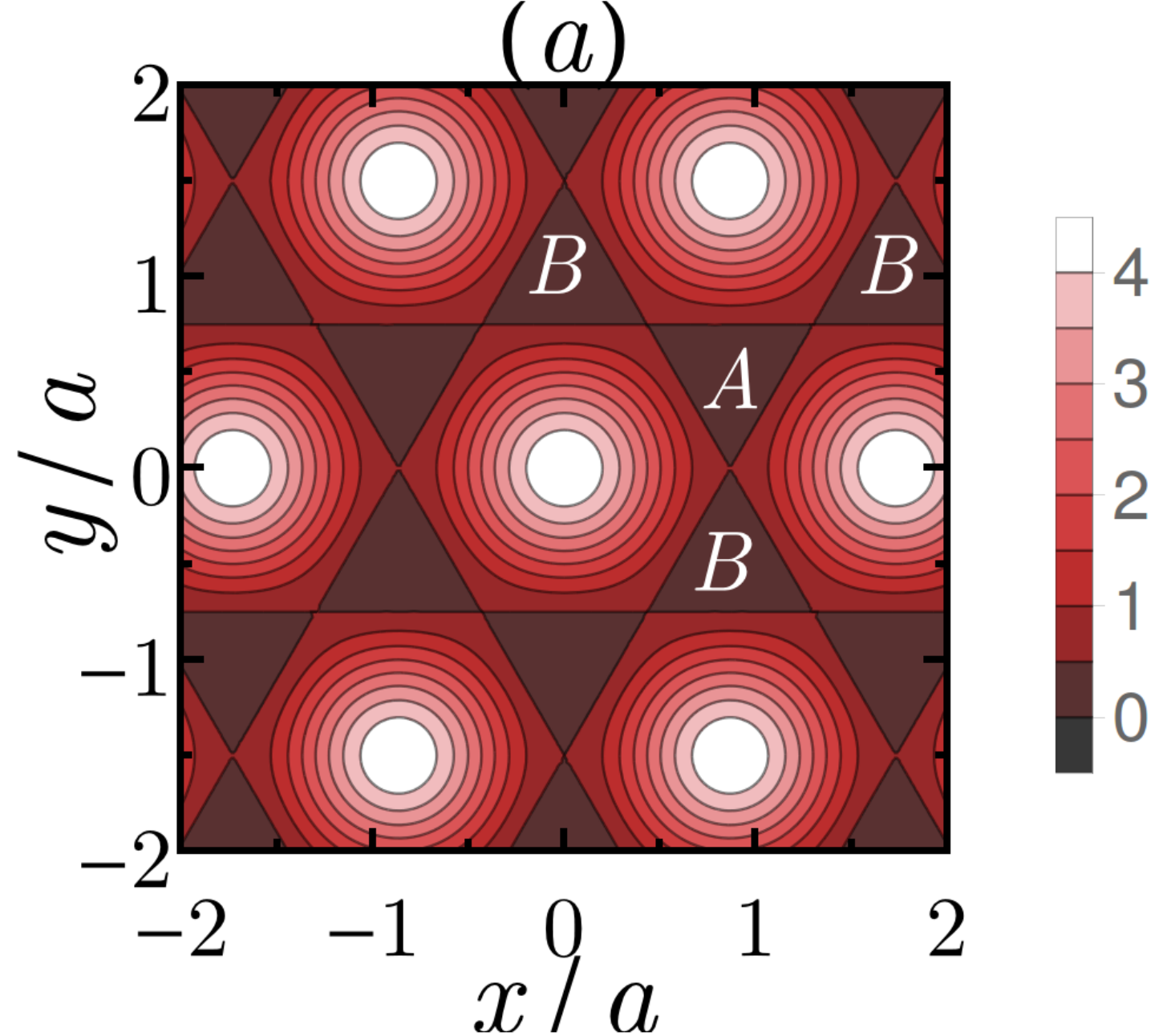}
\includegraphics[width =0.45\linewidth]{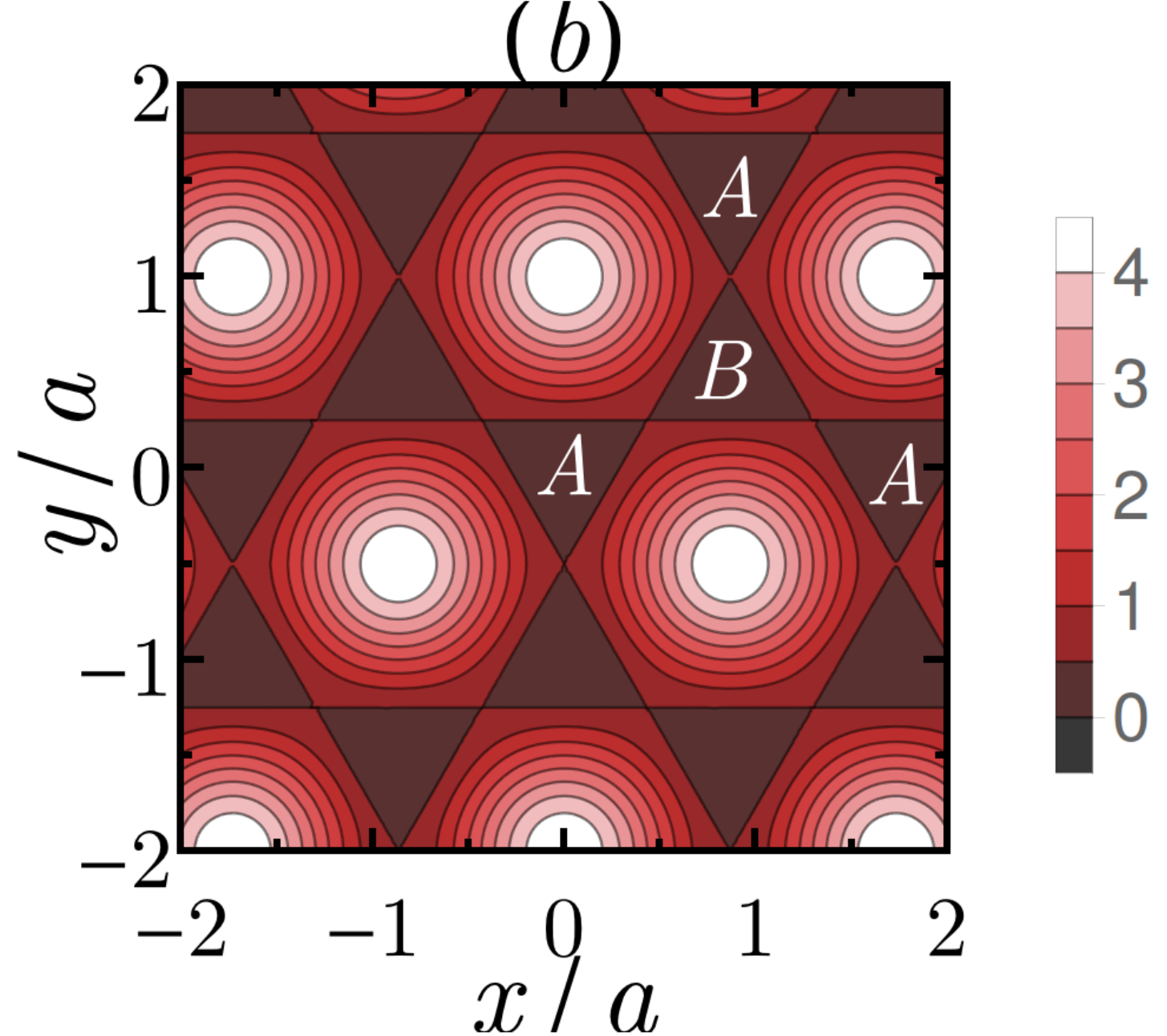}
\includegraphics[width =0.65\linewidth]{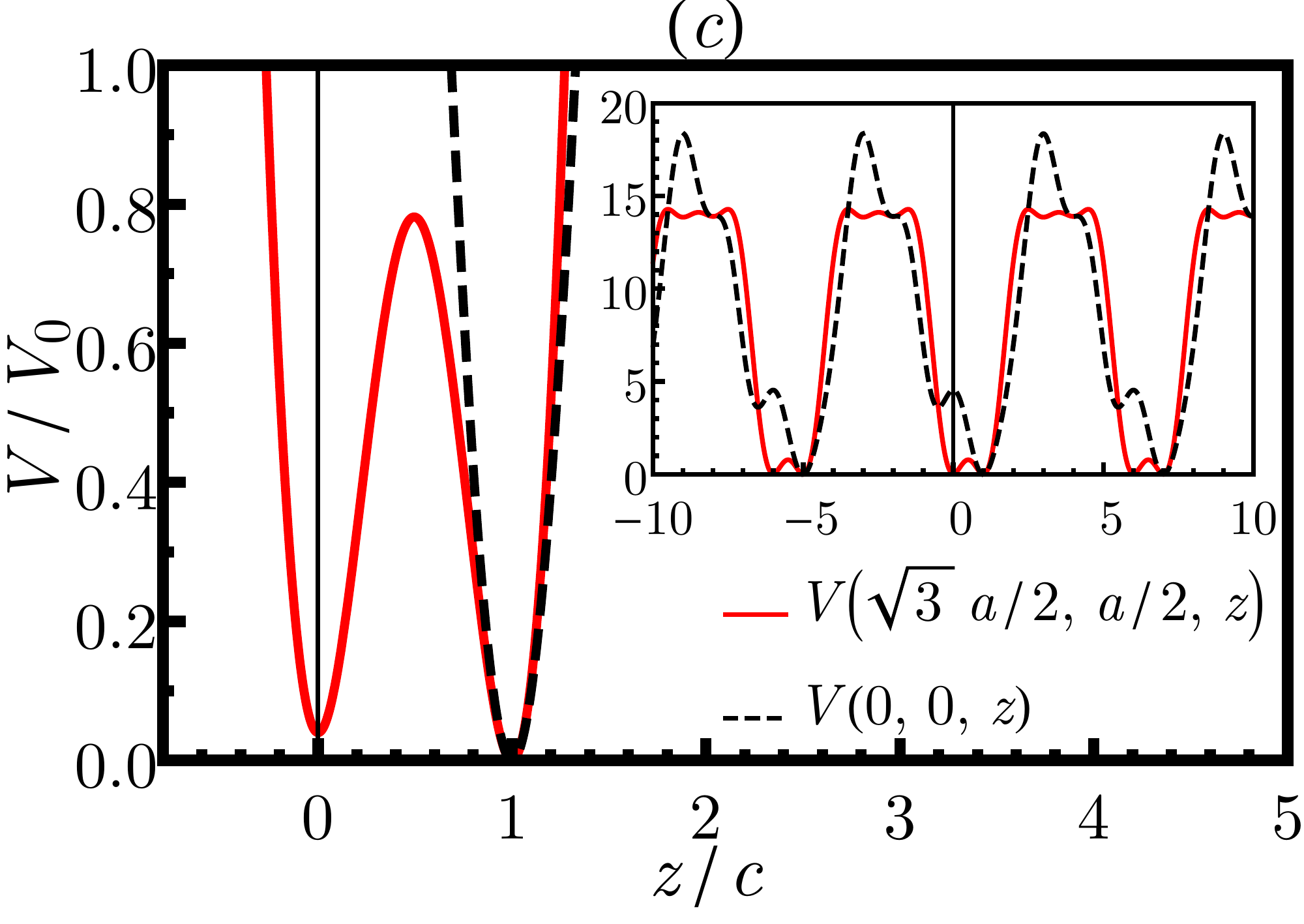}
\caption{ Color plots of the optical lattice potential (in units of $V_0$) in the $x-y$
  plane at (a) $z=0$ and (b) $z=c$. (c) The potential as a
  function of $z$  at $(\sqrt{3}/2 , 1/2)a $ (solid red line)
  showing A-B stacking and at $(0,0) $ (dashed black line) showing
  potential well in only one layer. Inset: The potential
  profile on a larger $z$ scale, showing the almost decoupled bilayers.}
\label{fig:1}
\end{figure}
we propose here a simple scheme of generating the Bernal stacked bilayer
honeycomb lattice using five lasers with the potential 
\bqa
 V(\mathbf{r})&=&V_0[
\cos (\mathbf{ p_1}\cdot\mathbf{ r})+\cos (\mathbf{ p_2}\cdot\mathbf{
  r})+\cos (\mathbf{ p_3}\cdot\mathbf{ r})+\frac{3}{2}]\\
\no & -\alpha&
V_0\left[\cos\left(\mathbf{ p_4}\cdot\mathbf{ r}-\phi_1\right)
-\frac{1}{6}
\cos\left(\mathbf{ p_5}\cdot\mathbf{ r}-\phi_2\right)-\frac{\sqrt{3}}{2}\right] 
\eqa 
where $\mathbf{p_{1}} = (\frac{2\pi}{\sqrt{3}a},\frac{2\pi}{3a},-\frac{2\pi}{3c})$,
$ \mathbf{p_{2}} =
(-\frac{2\pi}{\sqrt{3}a},\frac{2\pi}{3a},-\frac{2\pi}{3c})$ and 
$ \mathbf{p_{3}} = (0,-\frac{4\pi}{3a},\frac{4\pi}{3c})$ produces a
honeycomb lattice in the $x-y$ plane with lattice constant $a$, whose origin is shifting with
$z$ in such a way that it results in $A-B$ stacking with intralayer
distance c. These three layers however create a 2D pattern whose
origin is continuously shifting with $z$. In order to pick out the
correct $x-y$ planes corresponding to the bilayer, the
last two lasers, with $\mathbf{p_{4}} = (0,0,\frac{\pi}{3c})$,
$ \mathbf{p_{5}} =(0,0,\frac{\pi}{c})$, $\phi_1=\frac{\pi}{6}$ and
$\phi_2=\frac{\pi}{2}$, create a superlattice in the $z$
direction and picks up the planes $z=0$ and $z=c$, producing the
Bernal stacked bilayer. The full laser arrangement actually creates a
superlattice of bilayers, separated by a distance $6c$, but the large
distance between successive bilayers result in a system of decoupled
Bernal stacked bilayer honeycomb lattice. In addition, a bias $\Delta $ can be created between
the planes either by a trapping potential or
by application of a weak laser with wavelength $2c$.

The potential in the $x-y$
plane at $z=0$ and $z=c$ are plotted in Fig~\ref{fig:1} (a)
and (b), which shows the shifted honeycomb pattern in the two
layers. The potential as a function of $z$ is plotted in Fig \ref{fig:1}(c) for
in-plane location $(\sqrt{3},1)\frac{a}{2}$ (solid red line), where
$A$ and $B$ sublattices lie above each other, and
for the location $(0,0)$, where there is a potential well ($B$
sublattice) in only one plane (dashed black line). The inset shows the stack of decoupled
bilayers, repeated with period $6c$.

\subsection{ Biased bilayer and annular Fermi sea}
 
\begin{figure}[t]
\includegraphics[width =0.45\linewidth]{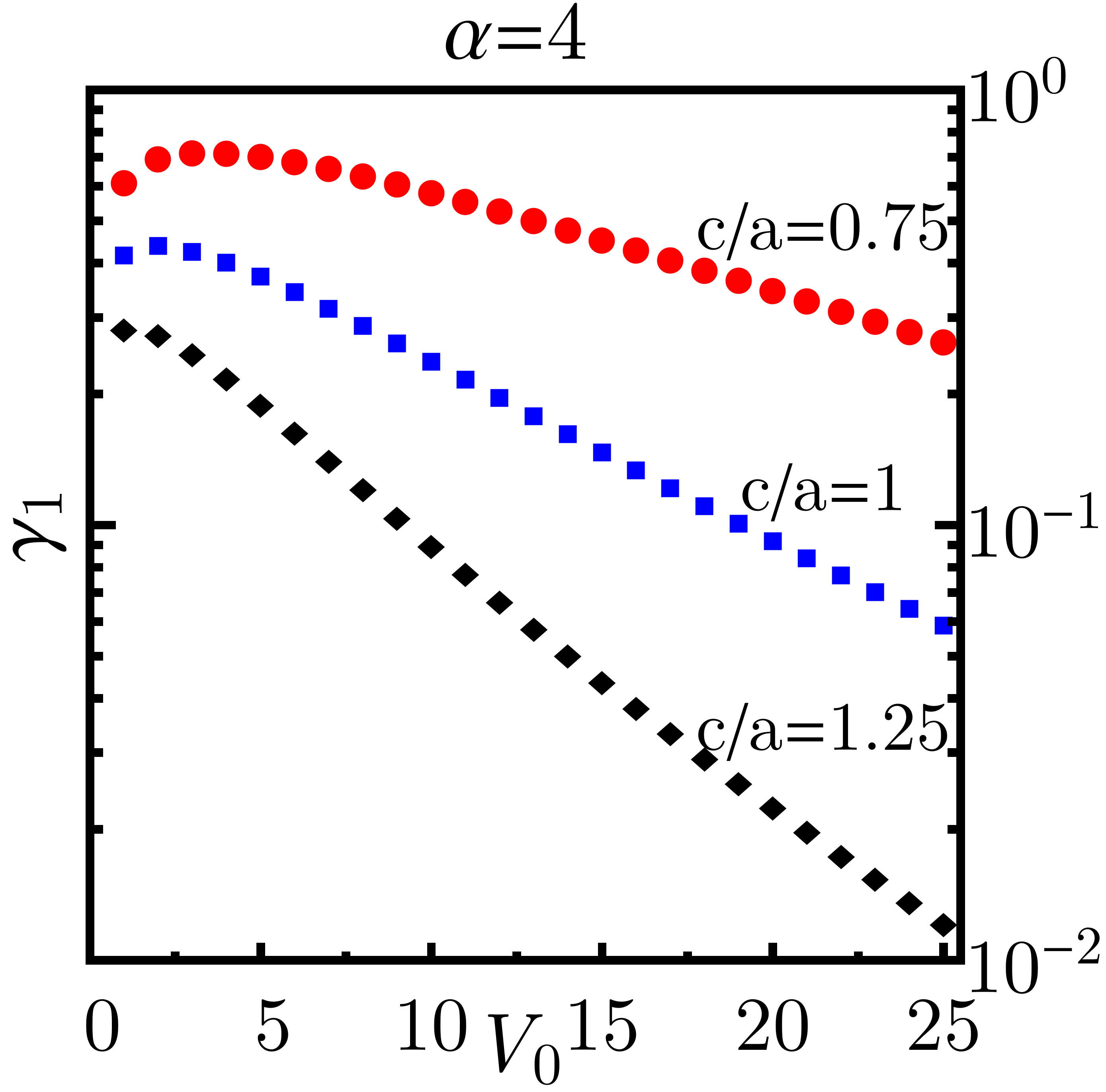}
\includegraphics[width =0.45\linewidth]{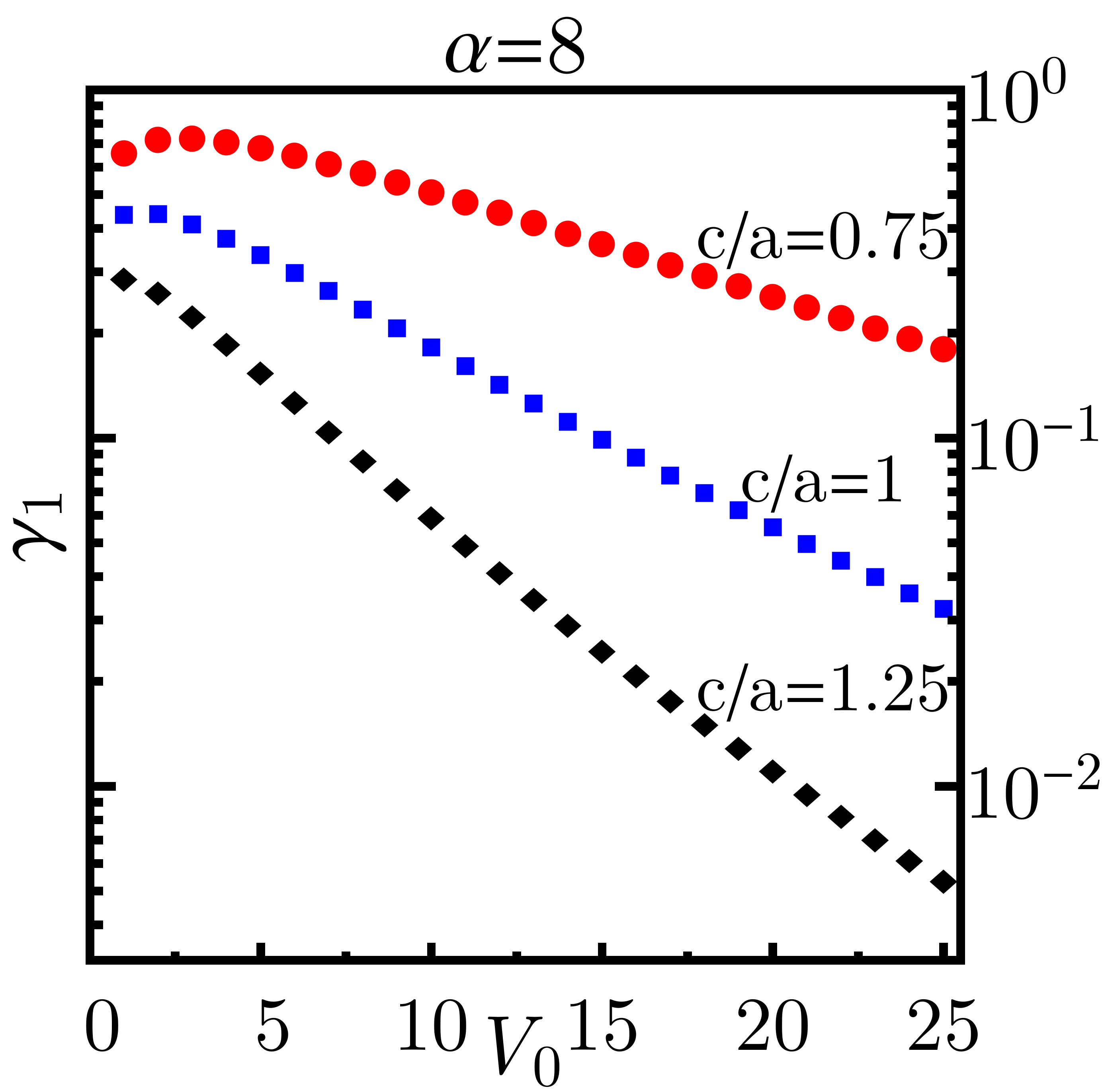}
\caption{The dependence of the interplanar coupling $\gamma_1$ on the
  lattice parameter $V_0$ for different values of $c/a$ for (a)
  $\alpha =4$ and (b)  $\alpha=8$.}
\label{fig:gamma1}
\end{figure}

The bilayer honeycomb lattice has four sites in its unit cell described by the sublattice (A
and B) and layer ($1$ and $2$) indices. The tight-binding
Hamiltonian~\cite{tight_binding} consists
of nearest neighbor in-plane tunneling $\gamma_0$ and an interlayer tunneling $\gamma_1$ between $A_1$ and
$B_2$ sites~\cite{footnote1}.
$\gamma_0$, which depends on $a$ and $V_0$, is known to follow~\cite{coldatom_bandparam_1,coldatom_bandparam_2},
$\gamma_0/E_R=1.16(V_0/2E_R)^{0.95}exp(-1.634\sqrt{V_0/2E_R})$~\cite{coldatom_bandparam_2,footnote2} (with $E_R=8\pi^2\hbar^2/27 ma^2$). 
$\gamma_1$ depends on $c$, $V_0$ and $\alpha$. Fig~\ref{fig:gamma1}
shows the variation of $\gamma_1$ obtained from a band structure
calculation for different values of $c/a$ and
$\alpha$~\cite{Wannier}. Since the parameters $c/a$ and $\alpha$ are
tunable in a cold atom setting, the ratio
$\gamma_1/\gamma_0$ can be tuned over a large range in the cold-atom
system. An independently tunable bias between the layers,
$\Delta$, can be added to gap out the single particle spectrum.

The tight binding band theory is written in the basis 
$\psi_k=(c_{A1k},c_{B1k},c_{A2k},c_{B2k})$ ~\cite{blg_review} as $H=\sum_k \psi^\dagger_k H_k\psi_k$ where,
\beq
H_k=\left(\begin{array}{cccc}%
\frac{\Delta}{2}& -\gamma_0f(\kk) & 0 & -\gamma_1\\
-\gamma_0f(\kk)^* & \frac{\Delta}{2} & 0 & 0 \\
0 & 0 & -\frac{\Delta}{2} & -\gamma_0f(\kk) \\
-\gamma_1 & 0 & -\gamma_0f(\kk)^* & -\frac{\Delta}{2}
\end{array}\right).
\eeq
and $ f(\kk)=e^{ik_ya/\sqrt{3}}+2e^{-ik_ya/2\sqrt{3}}\cos{k_xa/2}$.
$ \gamma_0 $ is the in-plane nearest
neighbour hopping,$ \gamma_1 $ is the c-axis hopping between the $A$
and $B$ lattice sites that lie on top of each other ($A1$ and $B2$)
and  $\Delta$ is the potential bias between the layer. Here $\gamma_0$
and $\gamma_1$ are much larger than $ \gamma_3 $ (hopping between $B1$
and $A2$),  and $\gamma_4$ (hopping between $A1$ and $A2$), which are
neglected in our calculation~\cite{blg2}. Near the Dirac points of
the in-plane Hamiltonian, the structure factor 
$ f(\mathbf{K}^{\pm}+\qq)\simeq -v_F(\pm q_x-iq_y) $,
 and hence the Hamiltonian in the two inequivalent valleys is
\beq
H_k=\left(\begin{array}{cccc}%
\frac{\Delta}{2}& v_Fke^{-i\theta_k} & 0 & -\gamma_1\\
v_Fke^{i\theta_k} & \frac{\Delta}{2} & 0 & 0 \\
0 & 0 & -\frac{\Delta}{2} & v_Fke^{-i\theta_k} \\
-\gamma_1 & 0 & v_Fke^{i\theta_k} & -\frac{\Delta}{2}
\end{array}\right).
\eeq
Here $v_F=\sqrt{3}a\gamma_0/2$ is the Fermi velocity and $\theta_k$ is
the azimuthal angle in the $\vec{k}$ space. The dispersion ~\cite{tight_binding} consists of four
bands 
\bqa
\pm E^{\sigma}_\kk&=&\pm\left[\Delta^2/4+\gamma_1^2/2+v_F^2k^2\right.\\
\no & & \left.+\sigma
  \sqrt{v_F^2k^2(\Delta^2+\gamma_1^2)+\gamma_1^4/4}\right]^{1/2}
\eqa
where $\sigma=\pm 1$.
The wavefunctions corresponding to
these bands are given by 
\bqa\label{eq:wavefn}
\phi_+^{\sigma}(k)&=&[u^\sigma_{\kk},v^\sigma_{\kk}e^{i\theta_{\kk}},w^\sigma_{\kk}e^{-i\theta_{\kk}},x^\sigma_{\kk}]\\ 
\no \phi_-^{\sigma}(k)& =&  [-x^\sigma_{\kk},w^\sigma_{\kk}e^{i\theta_{\kk}},-v^\sigma_{\kk}e^{-i\theta_{\kk}},u^\sigma_{\kk}] 
\eqa
where $\theta_k$ is the azimuthal angle in the momentum space, and  $
u^\sigma_{\kk} = \cos\chi^\sigma_{\kk}\cos\alpha^\sigma_{\kk}$, $
v^\sigma_{\kk}=\cos\chi^\sigma_{\kk}\sin\alpha^\sigma_{\kk} $, $
w^\sigma_{\kk}=\sin\chi^\sigma_{\kk}\sin\beta^\sigma_{\kk} $ and $
x^\sigma_{\kk}=\sin\chi^\sigma_{\kk}\cos\beta^\sigma_{\kk} $, with 
\bqa
\alpha^\sigma_{\kk}&=&\tan^{-1}\frac{v_{F}k}{E^\sigma_{\kk}-\Delta/2}\\
 \no \beta^\sigma_{\kk}&=&\tan^{-1}\frac{v_{F}k}{E^\sigma_{\kk}+\Delta/2}\\
\no \chi^\sigma_{\kk}&=&\tan^{-1}\frac{(v_F^2k^2-(E^\sigma_{\kk}-\Delta/2)^2)\sqrt{(E^\sigma_{\kk}+\Delta/2)^2+v_F^2k^2}}{\gamma_1(E^\sigma_{\kk}+\Delta/2)\sqrt{(E^\sigma_{\kk}-\Delta/2)^2+v_F^2k^2}}.
\eqa
For small momenta around the Dirac points, the $\sigma=+1$ bands are
high energy bands, gapped out on a scale of
$(\Delta^2/4+\gamma_1^2)^{1/2}$ and we focus on the two low energy
bands ($\sigma=-1$). The bands have a sombrero like dispersion
with a local maxima at $k=0$ of height $\Delta/2$ and a band minimum
on a circle of radius
$k_0=\frac{\Delta}{2v_F}\left[(\Delta^2+2\gamma_1^2)/(\Delta^2+\gamma_1^2)\right]^{1/2}$
with the energy at the band bottom given by
$E_0=\frac{\Delta/2}{\sqrt{1+\Delta^2/\gamma_1^2}}$. 

If the
chemical potential $\mu$ can be tuned to lie within the well
of the mexican hat, of depth $V_{dip}= \frac{\Delta}{2}(1-(1+\Delta^2/\gamma_1^2)^{-1/2})$, the Fermi sea is annular in shape, with outer and inner radii given by $v_F^2k_{\pm}^2=\frac{\Delta^2}{4}+\mu^2\pm
\sqrt{\mu^2(\Delta^2+\gamma_1^2)-\frac{\gamma_1^2\Delta^2}{4}}$. 

\begin{figure}
\includegraphics[width =0.45\linewidth]{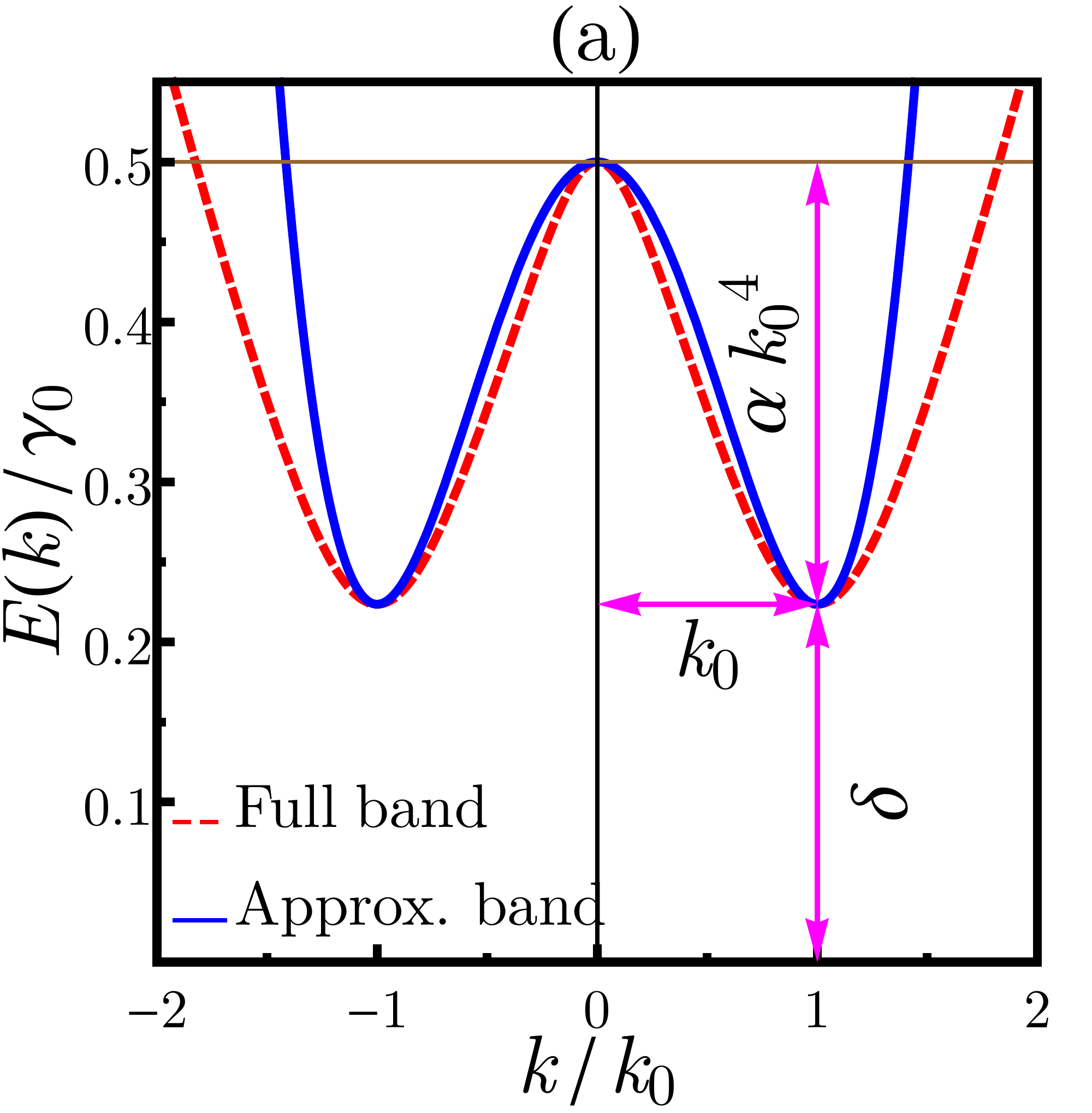}
\includegraphics[width =0.45\linewidth]{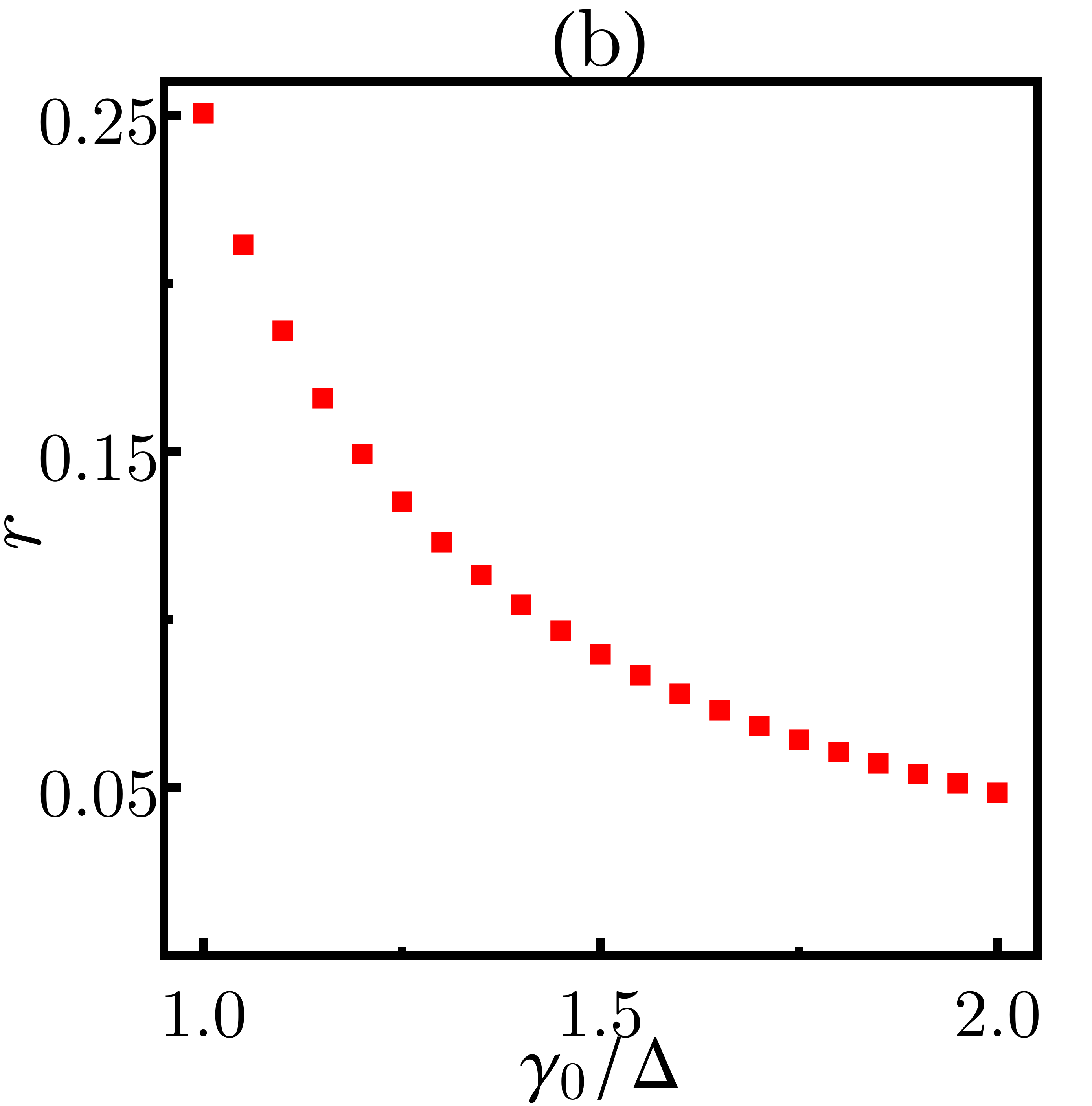}
\caption{(a)  The sombrero-like low energy dispersion of the actual
  band and the approximate band for $\gamma_1=\gamma_0=\Delta$. The
  figure shows how the parameters of the approximate band are obtained
  from the full dispersion(b) The ratio $r$ corresponding to the
  normalized density within the well of the sombrero of height
  $V_{dip}$ as a function of $\gamma_0/\Delta$. $\gamma_0/\Delta$
  corresponds to the largest $r\sim 0.25$}
\label{appfig:2}
\end{figure}

Near the Dirac points, the
low energy band dispersion takes the shape of a mexican hat, and one
can use 
an approximate quartic dispersion $
\epsilon_{k}=\epsilon_0(k^2-k_0^2)^2+\delta $ to reproduce the main
features of the system analytically. The parameter $ k_0 $ can
be chosen to coincide with the location of the band bottom, $ k_0=
\frac{\Delta\sqrt{\Delta^2+2\gamma_1^2}}{2v_F\sqrt{\Delta^2+\gamma_1^2}}
$. For the other two parameters the choice $
\epsilon_0=\frac{V_{dip}}{k_0^4} $ and $
\delta=\frac{\Delta}{2\sqrt{1+\Delta^2/\gamma_1^2}} $ matches the
dispersion at $k=0$ and $k=k_0$ and fixes the depth of the mexican
hat, as well as the energy at the band bottom.  Fig.~\ref{appfig:2}(a)
demonstrates a comparison between the detailed band dispersion and the 
approximate band dispersion in the relevant region of the Brillouin
zone for the parameter values $\gamma_0=\gamma_1=\Delta$. 

\subsection{Choice of band parameters}

For bilayer graphene the values of the tight binding parameters are
fixed, a standard profile being $ \gamma_1\simeq 0.4\text{eV} $, $
v_F=10^6\text{m/s} $ and a typical bias voltage $
\Delta\simeq10-100\text{meV} $~\cite{blg2}. However for the bilayer
honeycomb optical lattice there is a flexibility in choosing the
parameters of interest by tuning the laser profiles. 
The two main considerations which define the set of optimal parameters
are (i) the depth of the Mexican hat potential should be as large as
possible within other experimental constraints. This provides a large
temperature window over which the effects of the ring-shaped Fermi sea
can be seen clearly. (ii) the width of the Mexican hat dip (in
momentum space) should be as large as possible. A wide well implies
that a large density change corresponds to
a small chemical potential change and hence the acceptable error within which
densities need to be fixed in experiments increases.  

The phenomena associated with annular Fermi sea would be easier to 
observe if the well of the sombrero is deeper (less affected by
thermal fluctuations) and wider (a large change in the experimentally
controlled density results in a relatively small change in $ \mu
$). While this implies large $\Delta/\gamma_1$ and large
$\Delta/\gamma_0$ respectively, $\Delta/\gamma_1$ is bounded by the
constraint that the bias cannot be larger than the c-axis
bandwidth. $\Delta/\gamma_0$ is constrained by the fact that for
$\Delta/\gamma_0\gg 1$, the dispersion along the line connecting the
two valleys is very flat and the description in terms of populating
the individual sombreros around the different valleys do not remain valid. We define $r$ to be 
the ratio of the density at which the annulus shaped Fermi surface disappears, to the area of 
the Brillouin zone. We find that $r$ is controlled by
$\gamma_0/\Delta$ and keeps increasing with decreasing $\gamma_0/\Delta$, 
as shown in Fig.~\ref{appfig:2}(b). The tradeoff is optimized for
$\Delta=\gamma_0=\gamma_1$ which corresponds to $r\sim 0.25$.
We note that the independent tunability of $V_0$, $a$, $c$ and $\alpha$
provides a wide latitude in choosing experimental parameters.


The Van der Waals interaction between the fermions, described by a
scattering length $a_s$, leads to a local Hubbard interaction $H_{int}=U\sum_{i\tau} n_{i\tau\up}n_{i\tau\dn}$, where
$\tau$ denotes
the sublattice and layer indices.  In the deep
lattice limit, $U=\frac{4\pi a_s\hbar^2
}{27a^2cm}\left(\frac{2\pi^2V_0}{E_R}\right)^{3/4}  \sqrt{\frac{\alpha}{\sqrt{3}}+8}$,
and it can be tuned by changing $a_s$, $V_0$ or $\alpha$. This can be
used, for example, to quench the system across the critical
interaction strength corresponding to the Stoner instability, which we
will discuss in the later part of this paper.

Having shown the wide range of available experimental parameters, we now focus on the signatures of the
sombrero type dispersion, especially of the presence of two Fermi
surfaces, in the static susceptibility of this
system.

\section{\label{sec:pol}Response in Non-Interacting System}

The response of the fermions to potential perturbations is governed by the susceptibility 
\beq
\Pi^0(\qq,\omega)=-\sum_{ss'\kk} \frac{n_F(sE_\kk)-n_F(s'E_{\kk+\qq})}{\omega+sE_\kk-s'E_{\kk+\qq}}F^{ss'}(\kk,\qq)
\eeq
with 
$F^{ss'}(\kk,\qq)=|\phi_s^\dagger(\kk)\phi_{s'}(\kk+\qq)|^2$,
where $\phi_s(k)$ is the band eigenfunction. It is then clear that the
wavefunction overlaps $F^{ss'}(\kk,\kk+\qq)$ can be written in terms of the functions $ u_\kk $,$ v_\kk $,$ w_\kk $ and $ x_\kk $ defined above,
\begin{widetext}
\bqa
F^{++}(\kk,\kk')&=(uu'+xx')^2+v^2v'^2+w^2w'^2+2(uu'+xx')(vv'+ww')\cos\Delta\theta+2vv'ww'\cos2\Delta\theta \nonumber \\
F^{+-}(\kk,\kk')&=(xu'-ux')^2+v^2w'^2+w^2v'^2+2(xu'-ux')(vw'-wv')\cos\Delta\theta-2vv'ww'\cos2\Delta\theta
\eqa
\end{widetext}
where we have used a shortened notation $\kk'= \kk+\qq $, and the
angle between $\kk$ and $\kk+\qq$, $\Delta \theta$,
is given by $ \cos\Delta\theta=\frac{k+q\cos\theta}{|\kk+\qq|} $,
$\theta$ being the angle between $\kk$ and $\qq$. The
index $\sigma=-1$ is dropped from the functions for notational
brevity. 
We note the presence of both $\cos \Delta\phi$, reminiscent
of the chiral factors of graphene and $\cos 2\Delta\phi$, reminiscent
of the chiral factors of bilayer graphene, in the expression for the
overlap. This implies that the backscattering is not completely suppressed in this
case as in graphene. However 
In the limit $\gamma_1\rightarrow 0$ we
recover the suppression of backscattering present in graphene, while in
the $\Delta \rightarrow 0$ limit we recover the enhancement of
backscattering as found in bilayer graphene, and by tuning parameters
we can smoothly go from one limit to the other.

\begin{figure}[t]
\includegraphics[width=0.4\linewidth]{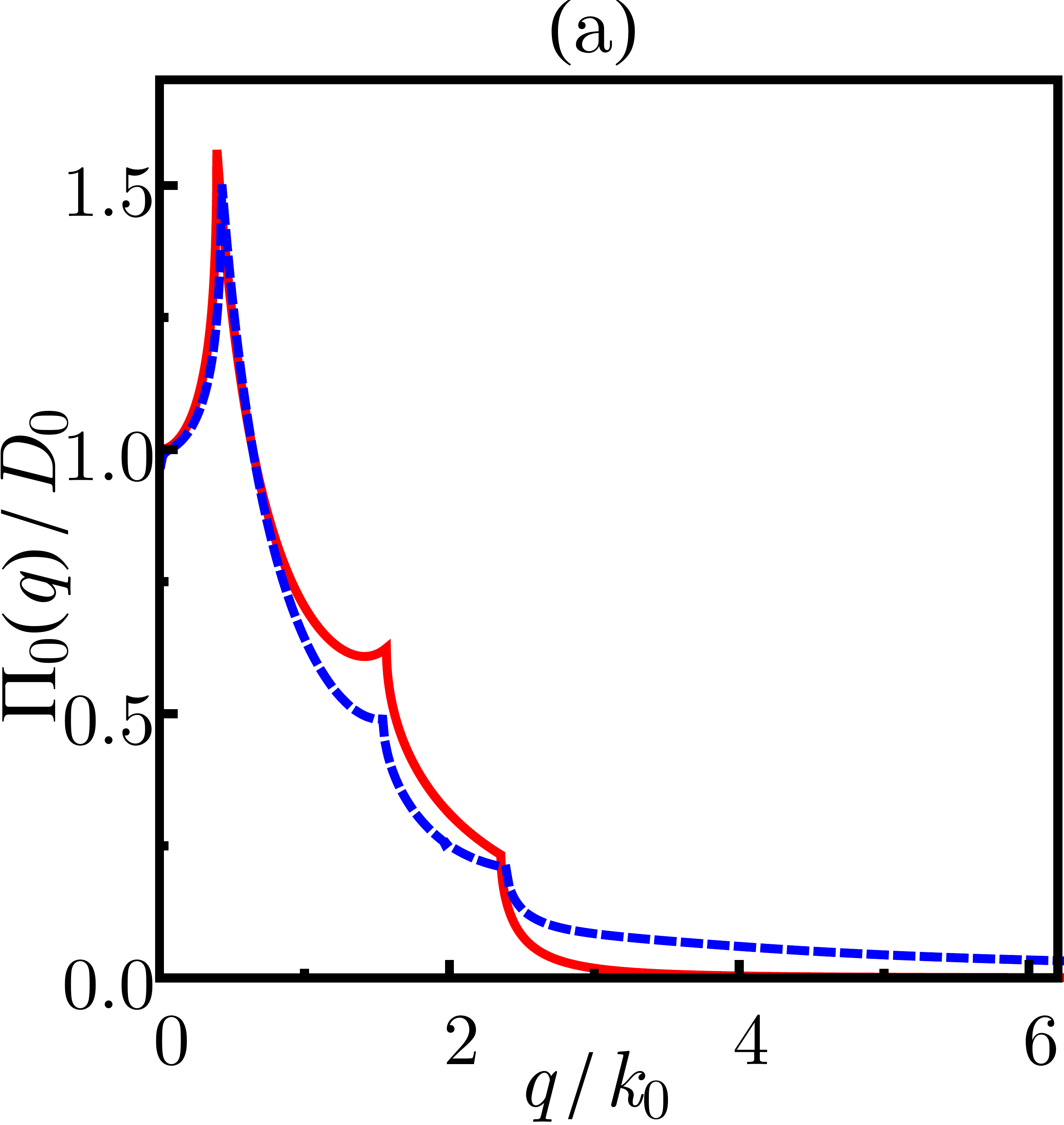}~~~~
\includegraphics[width=0.4\linewidth]{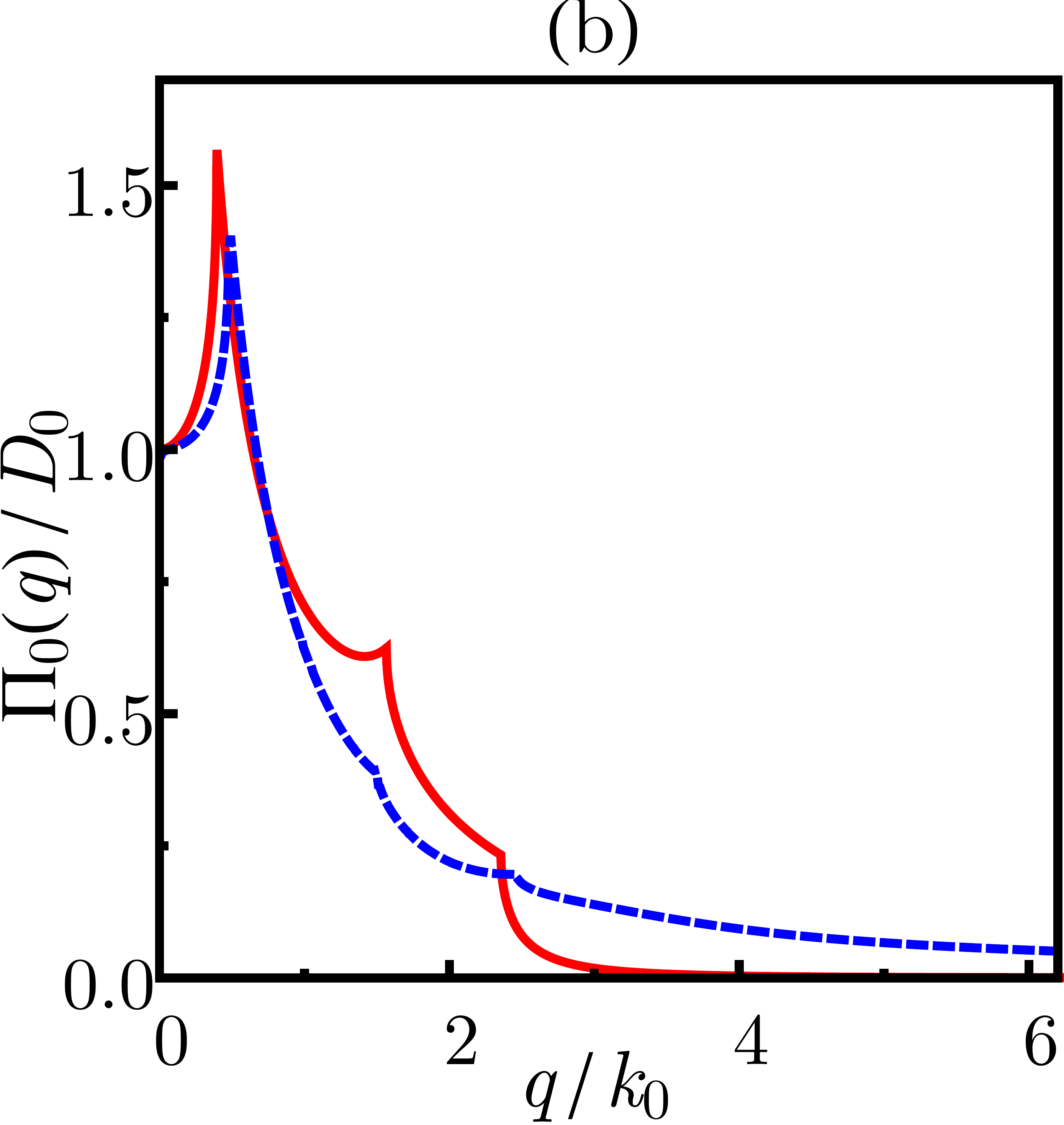}
\includegraphics[width=0.4\linewidth]{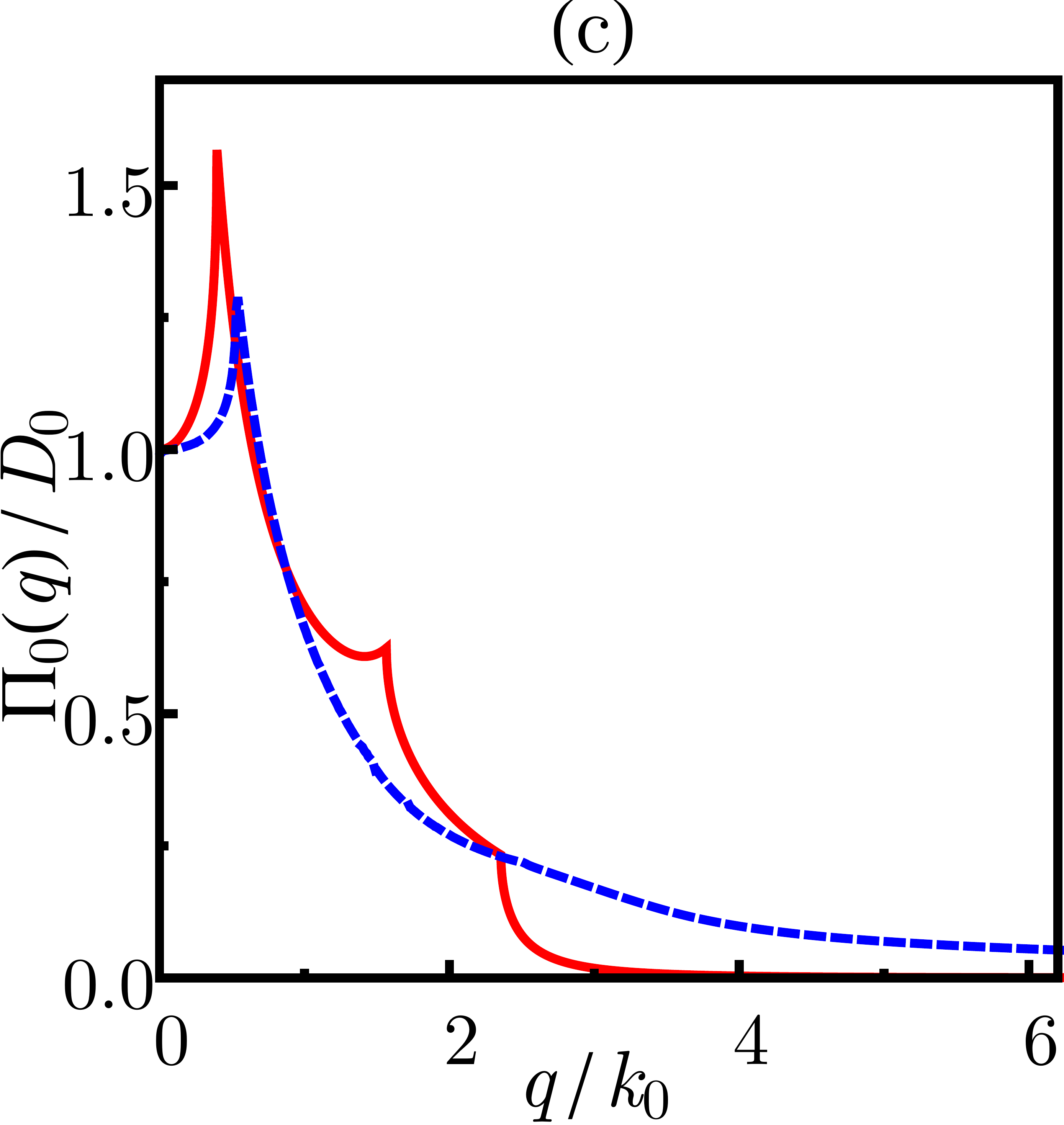}~~~~
\includegraphics[width =0.45\linewidth]{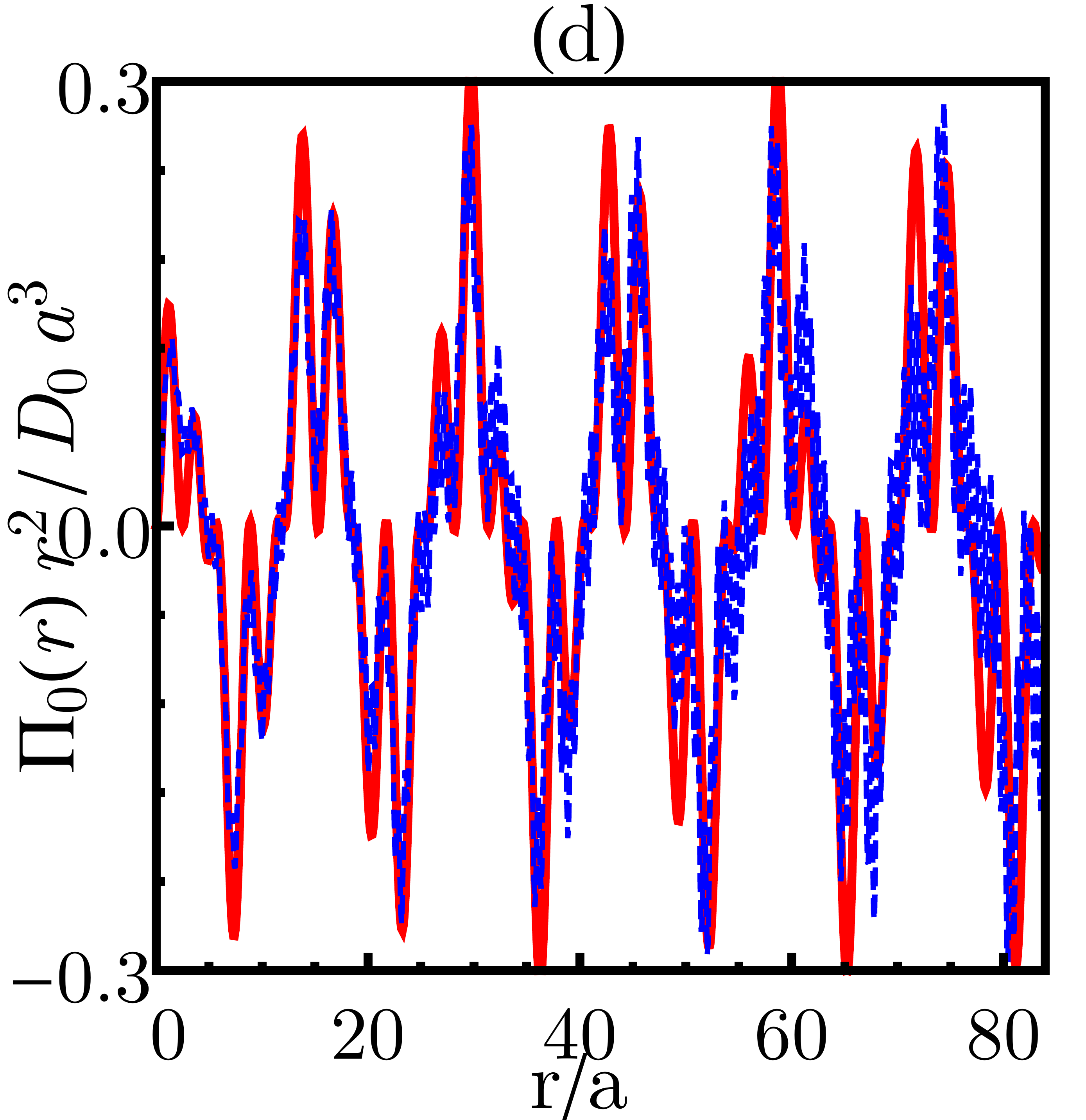}
\caption{The static susceptibility from the
  full band dispersion (dashed blue line) and the approximate
  dispersion (solid red line) for (a) $\gamma_1=2\Delta$ (b)
  $\gamma_1=\Delta$ and (c) $\gamma_1=\Delta/2$, normalized to the
  density of states at the Fermi level, $D_0$.(d) Friedel oscillations
  in real space for $\gamma_1=2\Delta$ showing beating
  patterns. $\mu=0.15 V_{dip}$ for all figures.}
\label{fig:2}
\end{figure}

The static susceptibility at $T=0$, $\Pi^0(\qq,0)$, is plotted as a function of $q$ in Fig. \ref{fig:2} (a-c) for decreasing values of
$\gamma_1$ (blue dashed line), with $\mu=0.15V_{dip}$. $\Pi_0(q)$ has
derivative singularities at $q=\delta k=k_+-k_-$ and $q=2k_\pm$, with
the strongest singularity occurring at $q=\delta k$. The $2k_\pm$
singularities weaken as $\gamma_1$ decreases and the system approaches
the limit of two decoupled graphene layers, where the absence of
backscattering leads to a weaker singularity ~\cite{Euyhyeon}.  The singularities in the susceptibility arise from
phase space restrictions whenever the Fermi surface shifted by the
vector $\qq$ touches the original Fermi surface. At $q=k_++k_-$, the
inner ring of the shifted Fermi surface touches the outer ring of the
original Fermi surface, while the outer ring of the shifted Fermi
surface touches the inner ring of the original Fermi surface. The
singular contributions from these cancel each other and hence, although backscattering connects fermions on the
outer Fermi surface to those on the inner Fermi surface at $q=k_++k_-$,  there
is no singularity at this wave vector. A similar situation arises when
$q=\delta k$, but in this case the singularities reinforce each
other. To obtain analytic insight into the behaviour of the
susceptibility, we work with one band model $\epsilon_k= \epsilon_0(k^2-k_0^2)^2+\delta$, where
$\epsilon_0$ and $\delta$ are chosen to match the low energy dispersion of the
actual conduction band. 
The static susceptibility within this approximation is given by
\beq
\Pi^0(\qq,i0^+)=-\sum_k \frac{n_F(\xi_k)-n_F(\xi_{k+q})}{\epsilon_k-\epsilon_{k+q}+i0^+}
\eeq
where $\xi_k=\epsilon_k-\mu$ and the $i0^+$ from the analytic
continuation is kept explicitly as it will play a vital role later.
At $T=0$, this can be reduced to 
\begin{widetext}
\beq
 \Pi^0(\qq,0^+)=
 -\frac{1}{2\pi^2\epsilon_0}\mathcal{R}\left[\int^{k_+}_{k_-}kdk\int^{\pi}_{-\pi}d\phi\left(\frac{1}{i0^++(k^2-k_0^2)^2-(k^2+2kq\cos\phi+q^2-k_0^2)^2}\right)\right]
\eeq
Rewriting the denominator as $ (2kq\cos\phi+x+y)(2kq\cos\phi+x-y) $,
with $ x=k^2+q^2-k_0^2$ and $y=\sqrt{(k^2-k_0^2)^2+i0^+}$, the angular integral gives
\bqa\label{eq:analyt_susc_int}
\Pi^0(\qq,0^+)&=&-\frac{1}{2\pi\epsilon_0}\mathcal{R}\left[\int^{k_+}_{k_-}kdk\frac{1}{y(k)}\left[\frac{1}{\sqrt{(k+q)^2-k_0^2+y(k)}\sqrt{(k-q)^2-k_0^2+y(k)}}\right.\right. \nonumber
\\ & &\left.\left. -\frac{1}{\sqrt{(k+q)^2-k_0^2-y(k)}\sqrt{(k-q)^2-k_0^2-y(k)}}\right]\right]
\eqa
\end{widetext}

After careful analytic
continuation, $ \frac{1}{\sqrt{(k^2-k_0^2)^2+i0^+}} $ can be replaced
by $ {\cal
  P}\left[\frac{1}{|k^2-k_0^2|}\right]-i\frac{\pi}{2}\delta(k^2-k_0^2)
$. Separating the real and imaginary parts of the two terms within the parenthesis of (\ref{eq:analyt_susc_int}) as $ A'(k) + i A''(k)$, we get, $\Pi^0(\qq,0^+)=-\frac{1}{2\pi\epsilon_0}\left[\mathcal{P}\int^{k_+}_{k_-}kdk\frac{1}{|k^2-k_0^2|}A'(k)+\frac{\pi}{4}A''(k_0)\right]  $.

The expressions for $ A'(k) $ and $ A''(k) $ are given by,
\begin{widetext}
\begin{equation}
\begin{aligned}
A'(k)&=\frac{\text{sgn}[\text{Re}(x+y)]\theta(\text{Re}(x+y)^2-4k^2q^2)}{\sqrt{\text{Re}(x+y)^2-4k^2q^2}}-\frac{\text{sgn}[\text{Re}(x-y)]\theta(\text{Re}(x-y)^2-4k^2q^2)}{\sqrt{\text{Re}(x-y)^2-4k^2q^2}} \text{ \ \text{and} \ }
\\ A''(k)&=\frac{\text{sgn}[\text{Im}(x+y)]\theta(4k^2q^2-\text{Re}(x+y)^2)}{\sqrt{4k^2q^2-\text{Re}(x+y)^2}}-\frac{\text{sgn}[\text{Im}(x-y)]\theta(4k^2q^2-\text{Re}(x-y)^2)}{\sqrt{4k^2q^2-\text{Re}(x-y)^2}}
\end{aligned}
\end{equation}
\end{widetext}
We note that $ A'(k)\sim (k-k_0) $ as $ k\rightarrow k_0 $. Hence $ \mathcal{P}\int\frac{kdk}{|k^2-k_0^2|}A'(k) $ is not a singular integral despite the denominator. The $ k\rightarrow k_0 $ limit of $ A''(k) $ either tends to zero (if $ q>2k_0 $) or to a nonzero value (when $ q<2k_0 $). This value diverges as $ q\rightarrow 2k_0^- $. It can also be seen that $ \mathcal{P}\int\frac{kdk}{|k^2-k_0^2|}A'(k) $ diverges as $ q\rightarrow 2k_0^- $. 
However this divergence is exactly canceled by the divergent piece coming from the part involving $A''(k)$. The susceptibility is thus smooth around $q\rightarrow 2k_0 $. After some algebra, the susceptibility function can be written as a simple quadrature

\begin{widetext}\
\beq
\Pi^0(\qq,0^+)=\frac{1}{2\pi\epsilon_0}\left[\frac{\pi}{2q}\frac{\Theta(2k_0-q)}{\sqrt{4k_0^2-q^2}}-{\cal P}\int^{k_+}_{k_-}
\frac{kdk}{k_0^2-k^2}\left(\frac{\Theta(q^2-4k^2)}{q\sqrt{q^2-4k^2}}-\frac{\text{sgn}[z]\Theta(z^2-4k^2q^2)}{\sqrt{z^2-4k^2q^2}}\right)\right]
\eeq
\end{widetext}
where $ z=2k^2-2k_0^2+q^2 $. This integral can be computed piecewise
for different ranges of $q$ and gives 
\bqa
\Pi^{0}(q) &=& \frac{1}{4\epsilon_0\pi q \lambda_q}f(q)\\
\no f(q)&=&\frac{\pi}{2}-\Theta(\delta k-q)\cos^{-1}\left(\frac{2q\lambda_q}{k_+^{2}-k_-^{2}}\right)
~~~~ 0< q< 2k_-\\
\no &=&\frac{\pi}{2}-\tan^{-1}\frac{\nu^-_q}{\lambda_q} ~~~~~~~~~~~~~~~~~~~~~~~~~~~~`
2k_-< q< 2k_0\\
\no &=&\tanh^{-1}\frac{\lambda_q}{\nu^-_q}-\Theta(q-2k_+)\tanh^{-1}\frac{\lambda_q}{\nu^+_q}
~~~~~~~ q>2k_0
\eqa 
where, $\lambda(q)=\sqrt{|k_0^2-q^2/4|}$ and $\nu^\pm(q)=\sqrt{k_\pm^2-q^2/4}$.

This analytic expression for susceptibility is also
plotted in Fig.\ref{fig:2} (a-c) with a solid red line. We see that the
susceptibility matches well for small $q\sim \delta k$, but the
weakening of the $2k_{\pm}$ singularities are not captured as the
wavefunction overlaps are left out in this approximation.  

The singularities of the susceptibility give rise to oscillations with corresponding wave-vectors in a)density patterns around defects (Friedel oscillation) and b) interaction between magnetic impurities (RKKY oscillations)~\cite{RKKY}. These
oscillations can be experimentally verified by creating localized defects in the system.
Since the oscillations asymptotically fall off as $ 1/r^2$, in Fig.\ref{fig:2} (d), we plot $r^2\Pi^0(r)$ for
$\gamma_1=2\Delta$ to illustrate the pattern of Friedel oscillations
in the system, clearly showing the wave vectors involved.

The non-trivial Friedel oscillations can be used as
  an experimental signature of the presence of two Fermi
surfaces in a system of non-interacting fermions. We next move on to
the consequences of having two Fermi surfaces for a system of fermions
interacting with repulsive Hubbard type interactions.

\section{ \label{sec:Stoner}Stoner Instability and incommensurate SDW}

We now consider a Bernal stacked bilayer honeycomb lattice of spin $1/2$ 
fermions which
are interacting with a repulsive on-site interaction
$U$. We note that we are considering a system of lattice fermions with
repulsive interactions close to half filling; hence the cold atom
system should be immune to the issues of atom loss which plague
continuum systems in a trap without an optical lattice. The presence
of the optical lattice reduces three body losses as long as the
lattice is deep enough to consider only one band per degree of freedom.
The presence of the lattice, however leads to Umklapp scattering and
hence to 2 body atom loss in the system. This is clearly an experimentally 
accessible regime, since
experiments with strong repulsive interactions in a lattice has
already been performed showing the presence of Mott
insulators~\cite{Bloch_review}. Typical experimentally measured timescales 
for atom loss, which can be achieved far from Feshbach
resonances are $\sim 1-2$ sec. in these
systems~\cite{doublondecay}. This relatively large timescale for atom
loss ensures that the phenomena described in the following sections
will not be washed out due to atom loss and should be easily
accessible to experiments on this system. 

\subsection{The Simple Stoner Picture}

The low energy theory of biased Bernal-stacked bilayer consists of
a sombrero like dispersion around the two inequivalent
Dirac points, $\vec{K}$ and $\vec{K'}$, which leads to a valley degree
of freedom, $\tau$.

\begin{figure}[t!]
\includegraphics[width=0.9\linewidth]{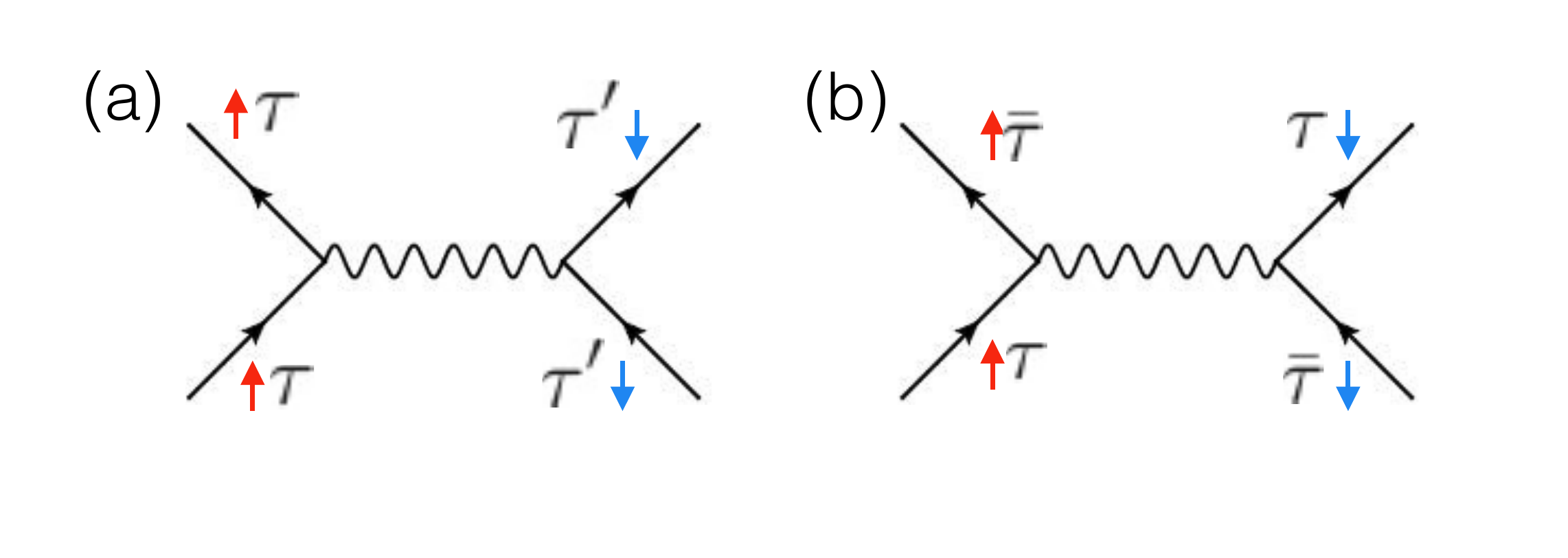}
\includegraphics[width=0.9\linewidth]{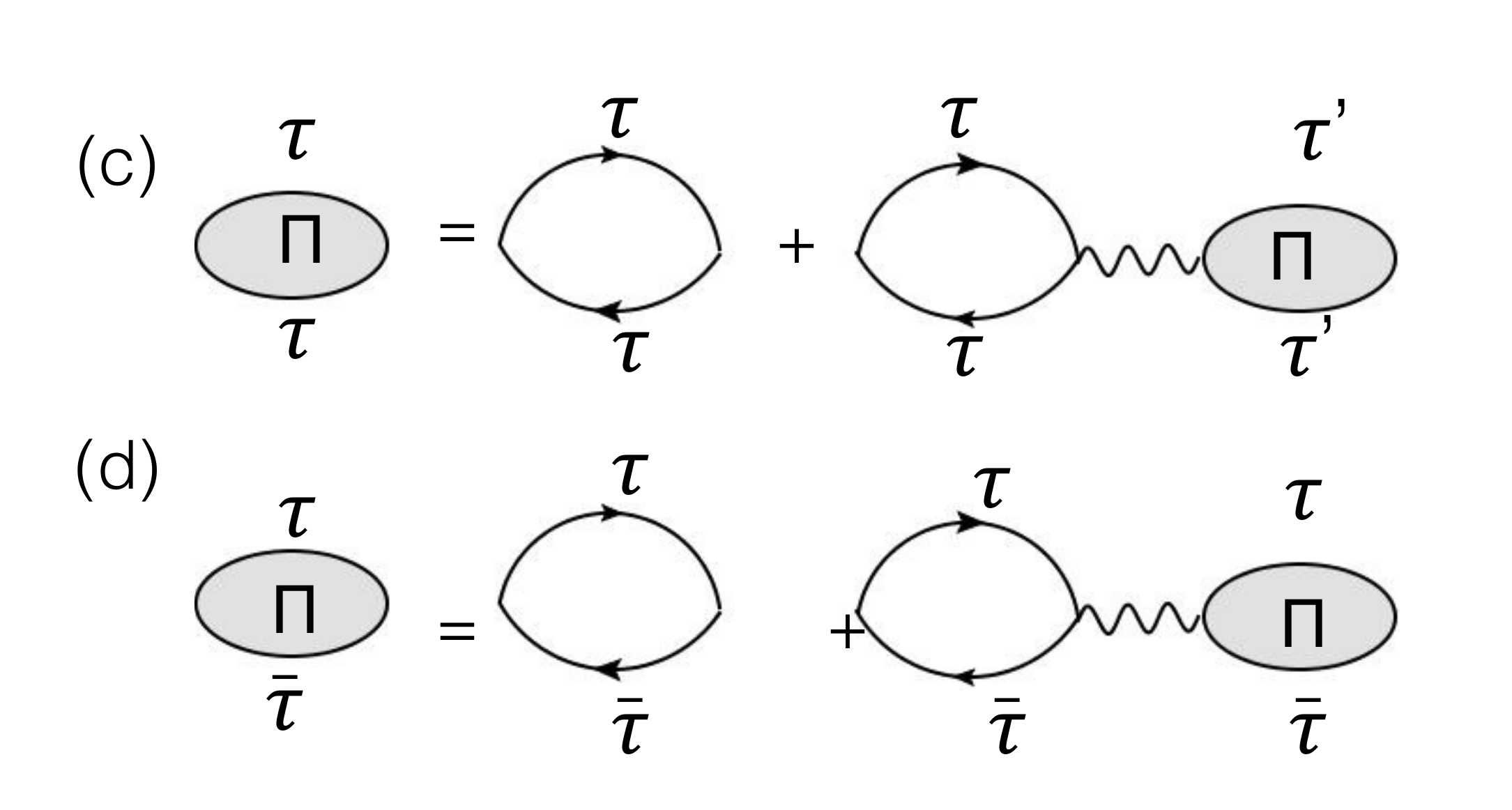}
\caption{(a) and (b): Bare interaction vertices showing spin and valley
  indices of the participating fermions. (c) and (d): RPA diagrams for
  the dressed susceptibility for intravalley (c) and intervalley (d)
  terms}
\label{appfig:3} 
\end{figure}

The local Hubbard interaction scatters fermions, both within a valley
and between valleys, as seen in the Feynman vertices of
Fig.~\ref{appfig:3}(a) \& (b). We assume that we are interested in
momentum transfers q much smaller than $|\vec{K}-\vec{K}'| $. Since we
will later be concerned with incommensurate SDW states with $q\sim
\delta k$, this is valid as long as $\delta k \ll |\vec{K}-\vec{K}'|$. This 
implies that if a fermion is scattered from $\vec{K}$ to $\vec{K}'$ valley, 
the other fermion must be scattered from $ \vec{K}'$ to $\vec{K}$ valley to 
conserve the total momentum. Consequently the $\tau $ indices of the Feynman 
vertex in Fig.~\ref{appfig:3}(b) has more restrictions than those in 
Fig.~\ref{appfig:3}(a). 

We use the notation $\Pi^0_{\tau\tau'}(q)$  to define the intra-valley and 
inter-valley bare susceptibility with $\tau$ as the valley index
($\tau=\pm1$ for $\vec{K}$ and $\vec{K'}$). We note that
$\Pi^0_{+-}(\vec{q})$ corresponds to a momentum transfer of
$\vec{q}+\vec{K}-\vec{K'}$. Within RPA approximation, we find that the 
intra-valley dressed susceptibility [Fig.~\ref{appfig:3}
(c)] is given by 
\beq
\Pi_{\tau\tau}(q)=\frac{\Pi^0_{\tau\tau}(q)}{1-U\sum_{\tau'}\Pi^0_{\tau'\tau'}(q)}=\frac{\Pi^0_{\tau\tau}(q)}{1-2U\Pi^0_{\tau\tau}(q)} 
\eeq
%
We note that unlike the spin-independent Coulomb interactions, the local Hubbard
interaction is only between fermions with opposite spins (see
Fig.~\ref{appfig:3}(a) and (b) ) and hence
there is no spin sum for the intermediate bubbles in the RPA
series. The spin states are fixed by the bare interaction vertex and
we omit them here for the sake of brevity. For the dressed intra-valley susceptibility,
the small momentum
transfer implies all the intermediate bubbles are of the intra-valley
type, but the fermions can be from either valley, resulting in the
factor of $2$ in the denominator, as shown in
Fig.~\ref{appfig:3}(c). In contrast, the inter-valley dressed [Fig.~\ref{appfig:3}(d)]
susceptibility is given by 
\beq
\Pi_{\tau\bar{\tau}}(q)=\frac{\Pi^0_{\tau\bar{\tau}}(q)}{1-U\Pi^0_{\bar{\tau}\tau}(q)}
\eeq
%
where the restriction on $\tau$ indices of bare vertices in \ref{appfig:3}(b) implies that there is no valley summation to be done and therefore there is no factor of 2 in the denominator. 

It is clear from the above argument that the intra-valley
susceptibility undergoes a Stoner instability at a critical strength
which is half that of the critical strength for the instability of the
inter-valley susceptibility. Thus within the Stoner/RPA approximation,
we can focus on the intra-valley susceptibility as the Stoner instability is
dominated by the intravalley scattering, with the Stoner criterion~\cite{Stoner}
\beq
1-2U\Pi^0({\bf q},0)=0
\eeq
where $2$ counts the valley degeneracy~\cite{footnote3}.
It is clear from Fig~\ref{fig:2} that, as $U$ is increased, the first
instability would occur at $q=\delta k$. From the analytic
calculations, $\Pi^0(\delta k,0) =(\pi/2) \Pi^0(0,0)$, and so for
$U > 1/2\Pi^0(\delta k,0)$, the spin fluctuation modes at
$q=k_+-k_-$ are most unstable, leading to an incommensurate spin-density
wave order.  As the
chemical potential is lowered towards the van-Hove singularity at the
band bottom, the
critical repulsion for the instability, $U_c$, goes to zero $\sim\mu^{1/2}$. The critical
coupling obtained from the simple Stoner criterion is plotted in
Fig.~\ref{fig:3} (a) and (b) for two different values of $\gamma_1/\Delta$, where the solid red curve corresponds to the
result obtained from the approximate band calculation, while the blue dashed
curve is obtained from the full band calculation. We note that
the two results track each other closely, showing that the
approximate band can be used to calculate the susceptibilities
accurately in this regime. This will be crucial in the next
section, where we will use the approximate band to include the effects of vertex corrections to the
Stoner instability.

\subsection{ Vertex Corrections and survival of Stoner Instability} 
\begin{figure*}[t]
\includegraphics[width=0.45\linewidth]{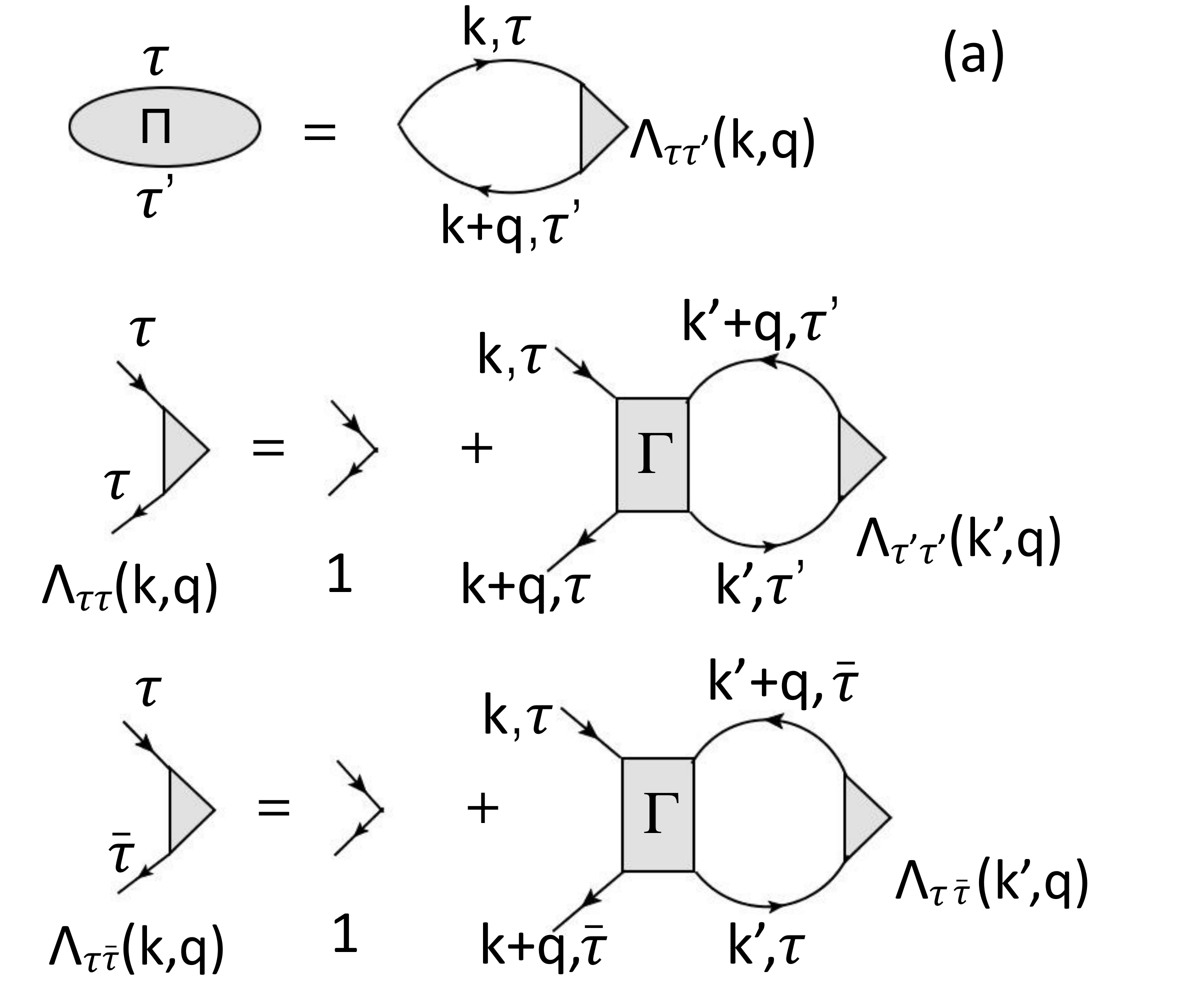}
\includegraphics[width=0.45\linewidth]{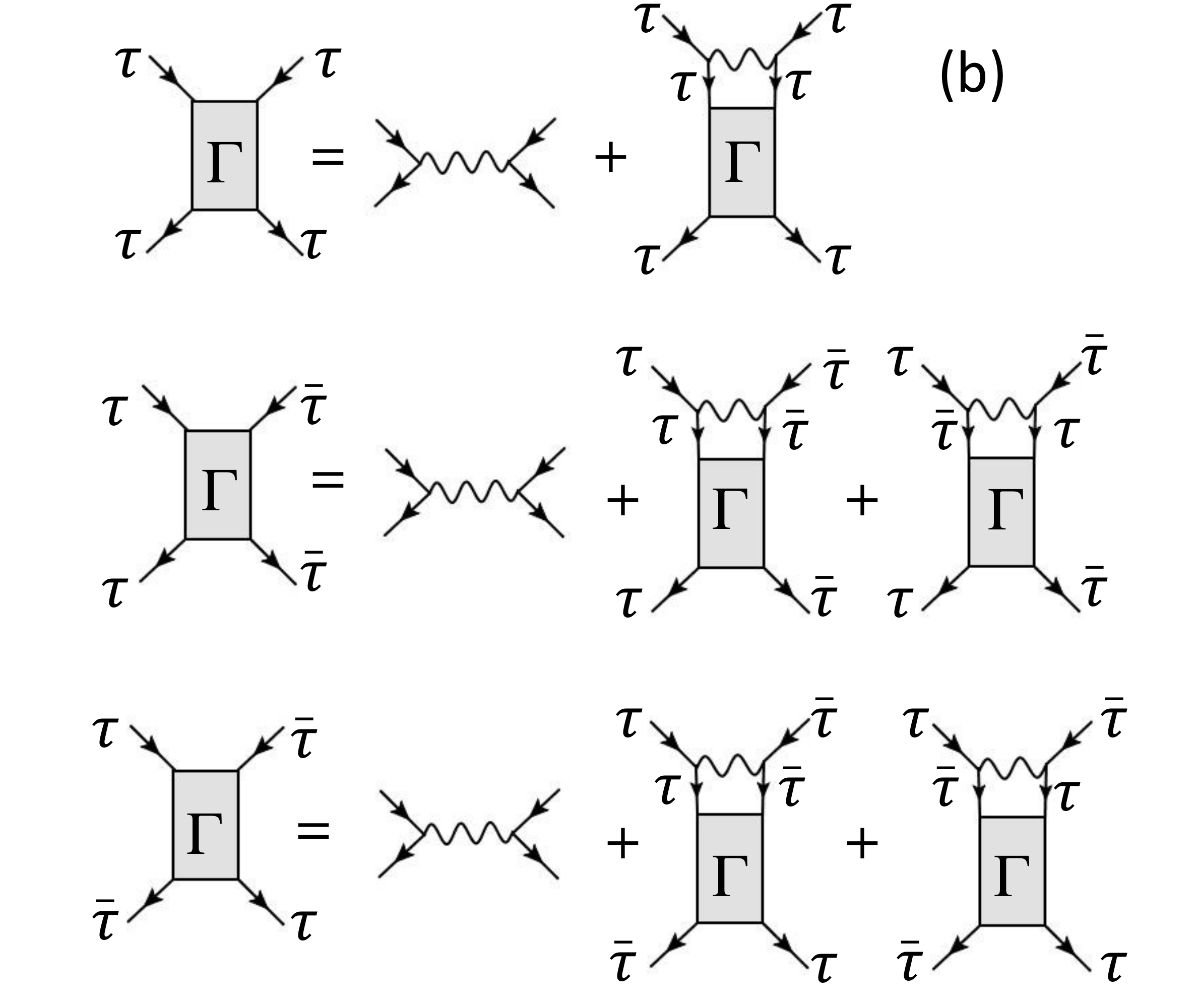}
\caption{(a) Feynman diagrams corresponding to vertex correction to
  susceptibility, $\Lambda$  and the integral equations for $\Lambda$
  from the partial resummation of the series. (b) The Feynman diagrams
  corresponding to ladder
resummation of the four-point vertex (effective interaction). The
$\tau=\pm 1$ indices correspond to the valleys $K$ and $K'$ }
\label{appfig:4} 
\end{figure*}

 Kanamori ~\cite{Kanamori} argued that since Stoner instability occurs at $U/W \sim 1$, where $W$ is
the bandwidth of the system, one should consider the manybody
renormalization of the interaction which can be strong enough to prevent the Stoner
instability. In this case, the instability occurs at weak
couplings as we move towards the band bottom, and survives the logarithmic
reduction of the repulsive interaction. To show this, we consider the vertex corrections to the spin
susceptibility within the analytic one-band model given by~\cite{Pekker}
\bqa
\label{eq:vertex1}
\Pi_{\tau\tau}(q)&=&\sum_{k} G_\tau(k)G_{\tau'}(k+q)\Lambda_{\tau\tau'}(k,q)
\eqa
where $k$ denotes both a momentum sum and a Matsubara frequency sum, and the greens function is $G(k)=G(\vec{k},\omega)=(\omega-\epsilon_k+\mu)^{-1}$. 
The Bethe-Salpeter type equation for the vertex function can be obtained from diagrammatic resummation [Feynman diagrams in Fig~.\ref{appfig:4}(a)],

\begin{widetext}
\bqa
\no \Lambda_{\tau\tau}(k,q)&=&1+\sum_{k',\tau'}\Gamma^{\tau\tau'}_{\tau\tau'}(k,k',q) G_{\tau'}(k')G_{\tau'}(k'+q)\Lambda_{\tau'\tau'}(k',q)
~~~ \text{and} \\
\Lambda_{\tau\bar{\tau}}(k,q)&=&1+\sum_{k'}\Gamma^{\tau\bar{\tau}}_{\bar{\tau}\tau}(k,k',q) G_{\bar{\tau}}(k')G_{\tau}(k'+q)\Lambda_{\tau\bar{\tau}}(k',q)
\eqa 
\end{widetext}
where we have assumed
$\Lambda_{\tau\bar{\tau}}=\Lambda_{\bar{\tau}\tau}$.The full four-point interaction vertex is approximated by the ladder sum leading to the T matrix
renormalization of the interaction strength.[Feynman diagram in Fig~.\ref{appfig:4}(b)]
\begin{widetext}
\bqa
\Gamma^{\tau\tau}_{\tau\tau} (k,k',q)&=&U-U\sum_{k_1}G_\tau(k_1)
G_\tau(-k_1+q)\Gamma^{\tau\tau}_{\tau\tau}(k_1,k',q) \\
\no \Gamma^{\tau\bar{\tau}}_{\tau\bar{\tau}} (k,k',q)&=&U-U\sum_{k_1}[ G_\tau(k_1)
G_{\bar{\tau}}(-k_1+q)\Gamma^{\tau\bar{\tau}}_{\tau\bar{\tau}}(k_1,k',q)+ G_{\bar{\tau}}(k_1)
G_{\tau}(-k_1+q)\Gamma^{\bar{\tau}\tau}_{\tau\bar{\tau}}(k_1,k',q)]
~~~ and \\
\no \Gamma^{\tau\bar{\tau}}_{\bar{\tau}\tau} (k,k',q)&=&U-U\sum_{k_1}[ G_\tau(k_1)
G_{\bar{\tau}}(-k_1+q)\Gamma^{\tau\bar{\tau}}_{\bar{\tau}\tau}(k_1,k',q) + G_{\bar{\tau}}(k_1)
G_{\tau}(-k_1+q)\Gamma^{\bar{\tau}\tau}_{\bar{\tau}\tau}(k_1,k',q)]
\eqa 
Since the Greens functions do not depend on $\tau$, we see that
$\Lambda_{\tau\tau}$ and $\Lambda_{\tau\bar{\tau}}$ are independent of
the index $\tau$ and hence obtain the T matrix renormalization of the
interaction vertex
\beq
\Gamma^{\tau\tau}_{\tau\tau} (q)=\left[\frac{1}{U}-C(q)\right]^{-1} ~~~~~~~~\Gamma^{\tau\bar{\tau}}_{\tau\bar{\tau}} (q)=\left[\frac{1}{U}-2C(q)\right]^{-1}  
~~~ and ~~~~~~~\Gamma^{\tau\bar{\tau}}_{\bar{\tau}\tau} (q)=\left[\frac{1}{U}-2C(q)\right]^{-1}  
\eeq 
\end{widetext}
where $C(q)=-\sum_k G(k)G(-k+q)$ is the non-interacting Cooperon propagator. Once
again it is clear that the intravalley susceptibility has stronger
instabilities. Focussing on the intravalley susceptibility, and dropping the $\tau$ indices, we get 
\begin{figure}[t]
\includegraphics[width =0.45\linewidth]{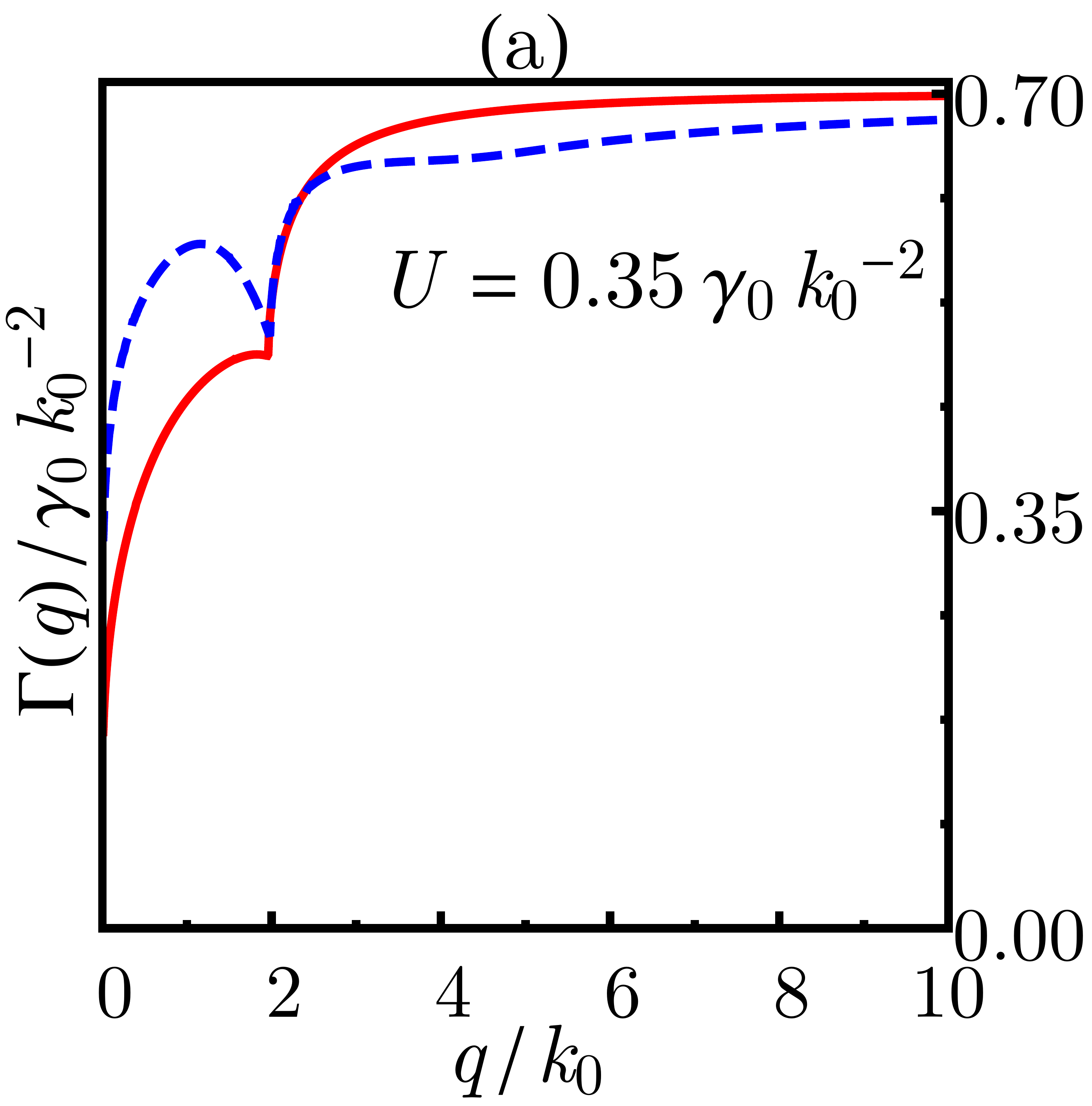}
\includegraphics[width =0.45\linewidth]{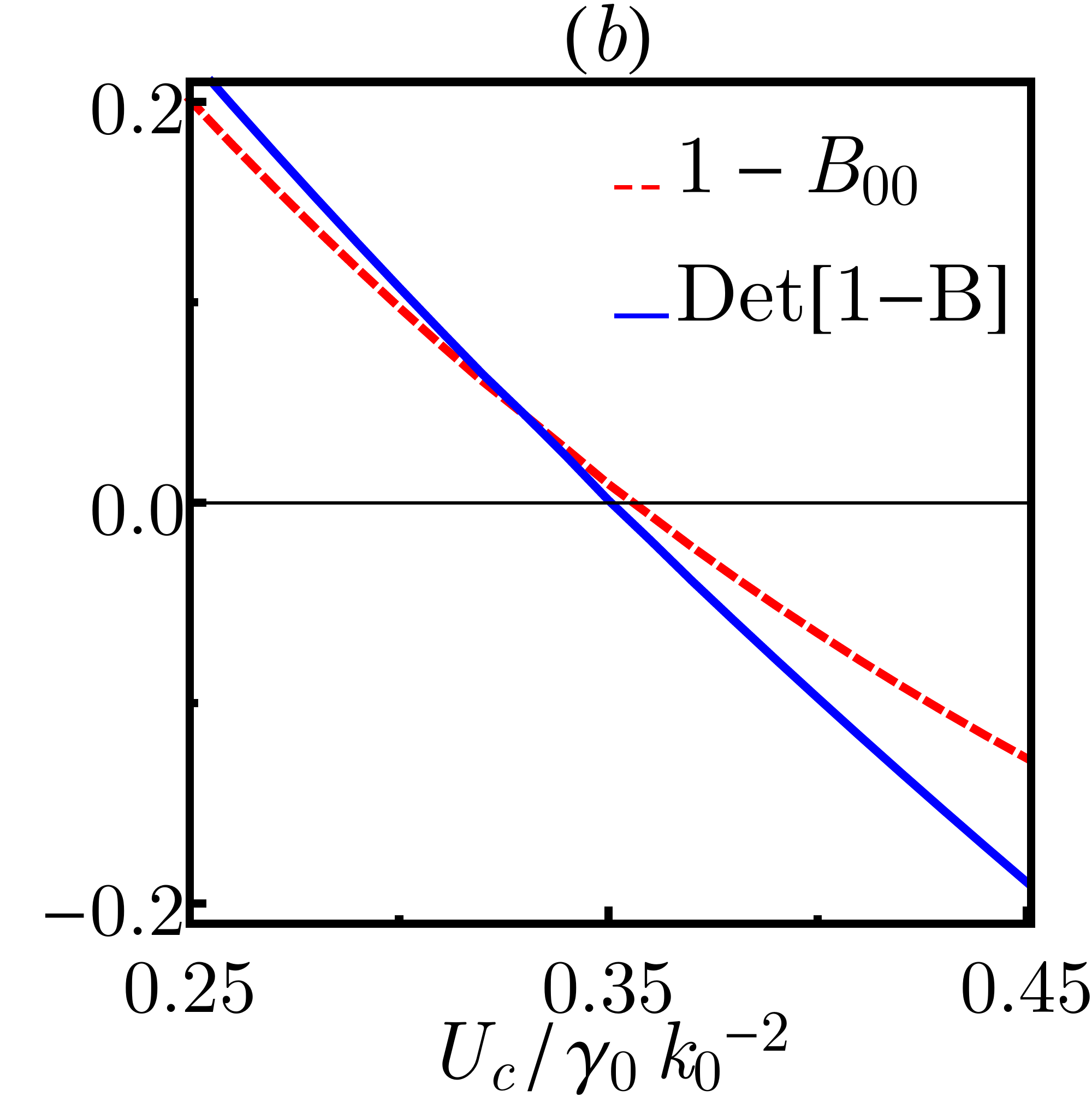}
\caption{(a)  The renormalized interaction $\Gamma(q)$ as a function
  of $q$ for $U=U_c=0.32 \gamma_0$ for the approximate one band (red solid line) and the full band (blue dashed line). (b)The
  comparison of $U_c$ obtained by setting $Det[1-B]=0$ (blue solid
  line)  and by setting $1-B_{00}=0$ (red dashed line). The two
  methods give comparable results. Both (a) and (b) are obtained for
  $\gamma_0=\gamma_1=\Delta$ and $\mu=0.15 V_{dip}$. }
\label{appfig:5}
\end{figure}
\bqa
\Pi(q)&=&\sum_{k} G(k)G(k+q)\Lambda(k,q)\\
\no \Lambda(k,q)&=&1+\sum_{k'}\Gamma(k+k'+q) G(k')G(k'+q)\Lambda(k',q)\\
\no \Gamma (q)&=&\left[\frac{1}{U}-C(q)\right]^{-1} +\left[\frac{1}{U}-2C(q)\right]^{-1}  
\eqa
Within the one band model, the Cooperon propagator is given by 
\beq
C(q)= \sum_k \frac{1-n_F(\xi_k)-n_F(\xi_{k+q})}{\omega-\xi_k-\xi_{k+q}}
\eeq
%
%
%
We note that for the approximate band, we can
take the momentum cutoff to $\infty$ without any ultraviolet
divergence, as the large $k$ dispersion $\sim k^4$ in this case. The
linear dispersion at large $k$ for the full band requires a cutoff of
$\sim 1/a$, while including the complete dispersion for the hexagonal lattice cuts
off the linear dispersion and gives a finite answer. We have checked by
explicit enumeration that in the range of $q$ that we are interested
in, the choice of different cutoffs do not affect our answers. A plot of $\Gamma(q)$ as a
function of $q$ for both the one-band and full-band model with $\gamma_0=\gamma_1=\Delta$ and
$\mu=0.15 V_{dip}$ at $U=U_c\sim0.32\gamma_0/k_0^2$, is shown in
Fig~\ref{appfig:5}(a). The effective interaction
increases with $q$ and the bare answer $2U$, with $2$ for valley
degeneracy factor, is recovered in the large $q$ limit.
 
The divergence of the susceptibility is governed by a divergence of
the vertex factor $\Lambda$~\cite{Pekker}, which is controlled by contributions from
the momentum space where the two poles of the Green's functions $G(k)$ and
$G(k+q)$ come close to each other. For the static susceptibility
(zero external frequency), the main
contribution comes when $\vec{k}$ and $\vec{k'}$ both lie on a Fermi
surface. There are two possibilities: $|\vec{k}|$ and $|\vec{k'}|$ can lie
on either of the two Fermi wave-vectors $k_+$ and $k_-$. We also
set $|\vec{q}|=\delta k$, the thickness of the Fermi sea, where the
strongest instability is seen in the RPA approximation. We have
checked by varying $|\vec{q}|$ that this choice corresponds to the
strongest instability. We also find that the choice
$|\vec{k}|=|\vec{k'}|=k_+$ corresponds to the strongest
instability. This can be seen from the fact that the effective
interaction $\Gamma(q)$ increases with $q$, i.e. the renormalization
becomes less effective with increasing $q$. 

So, we fix the value of $|\vec{k}|$ and $|\vec{k'}|$ in all the terms
of the integral equation for $\Lambda$, except the product of the
Green's functions, which are rapidly varying function of the radial
co-ordinates~\cite{Pekker}. The integral equation for $\Lambda$ then reduces to 
\beq
 \Lambda(\hat{k},\vec{q})=1+\int
 d\hat{k'}\Gamma(k_+(\hat{k}+\hat{k'})+\vec{q})
 \Lambda(\hat{k'},\vec{q}) I(\hat{k'},\vec{q})
\eeq
where
\bqa
 I(\hat{k'},\vec{q})&=&\frac{1}{4\pi^2}\int k'dk' G(k')G(k'+q) 
\eqa
%
Without loss of generality, we assume that $\vec{q}$ lies along the $x$
axis. The integral equation is then a function of two angles, $\theta$
corresponding to $\hat{k}$ and $\theta'$ corresponding to
$\hat{k'}$. Expanding $\Lambda(\theta)=\frac{1}{2\pi}\sum_m
\Lambda_me^{im\theta}$,  and constructing the vector $\hat{\Lambda}$ of the fourier
components, we can then write a matrix equation
\beq
\hat{\Lambda}=(1-B)^{-1}\hat{w} 
\eeq
where $\hat{w}_m=\delta_{m0}$ and 
\beq
B_{mn}=\frac{1}{2\pi} \int_0^{2\pi} d \theta' \Gamma(\theta,\theta')I(\theta')e^{in\theta'-im\theta} 
\eeq
The instability criterion is then given by 
\beq
Det(1-B)=0
\eeq
However a simpler criterion $1-B_{00}=0$ gives an $ U_c$ which is remarkably close to the full answer
[Fig.~\ref{appfig:5}(b)],
which shows that the instability is essentially driven by the $m=0$
mode. We find that
increasing the size of the matrix $B$ does not lead to appreciable
changes, thus confirming that higher $m$ modes do not contribute to
the instability.

\begin{figure}[t]
\includegraphics[width =0.45\linewidth]{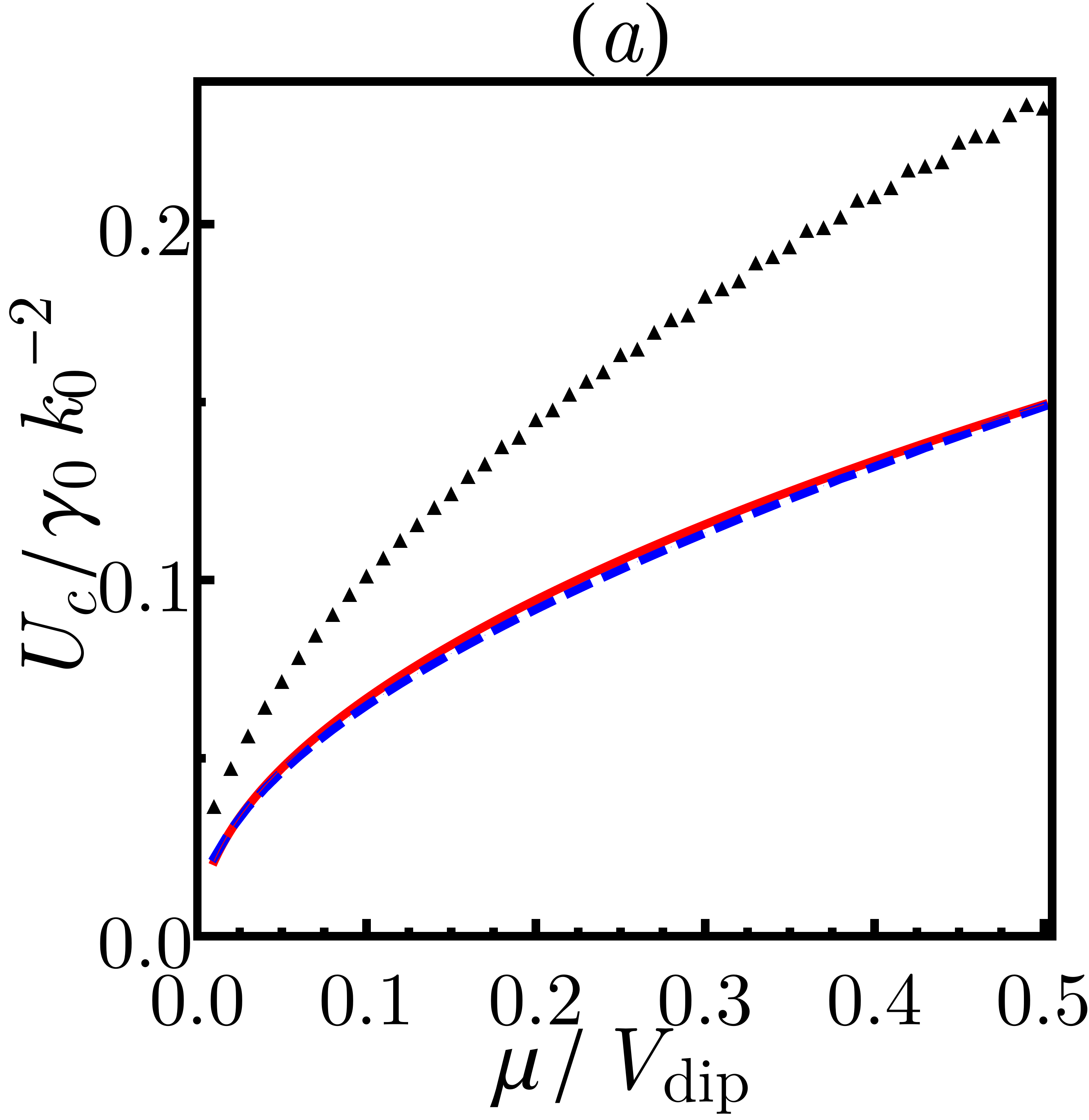}
\includegraphics[width =0.45\linewidth]{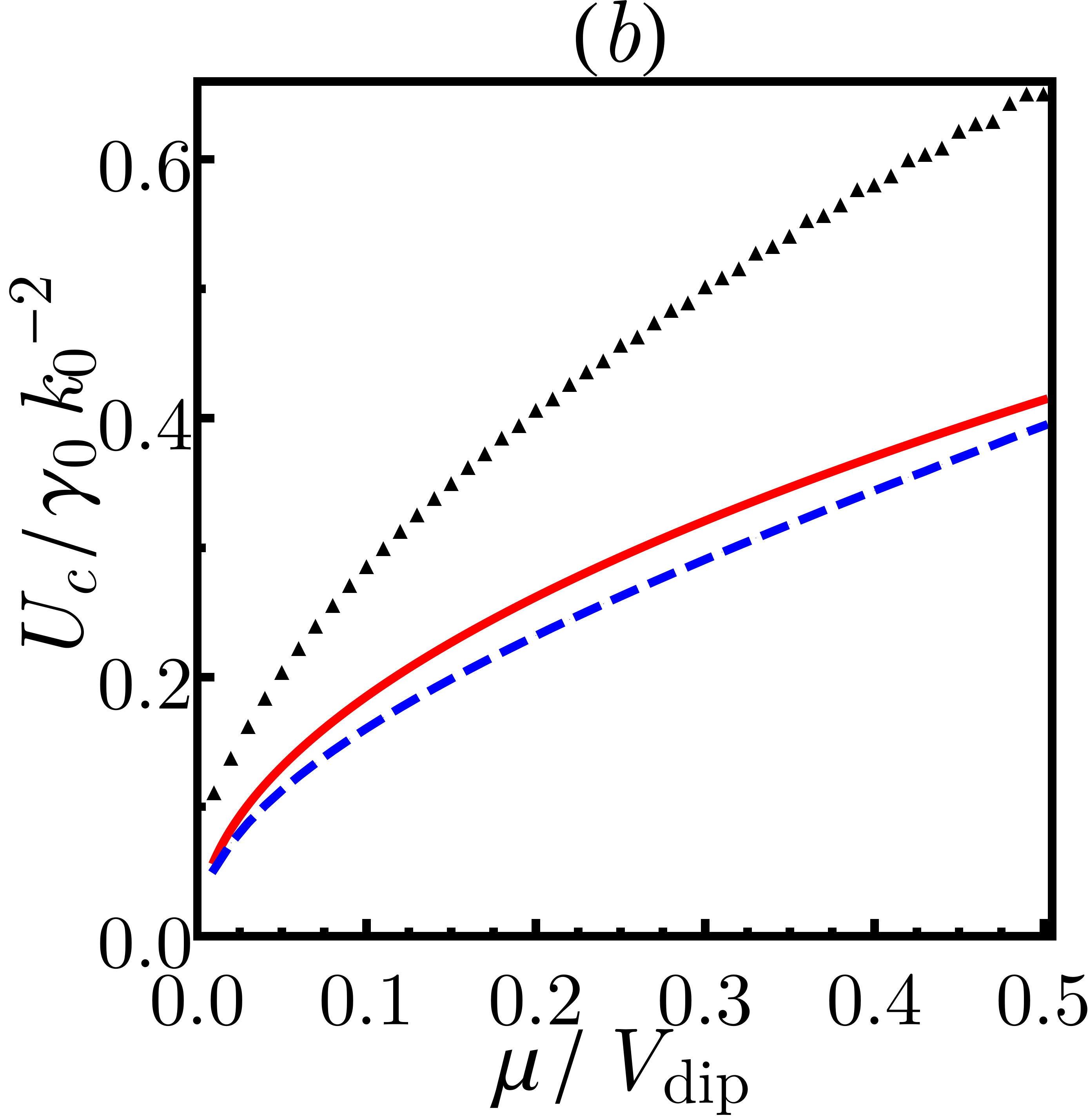}
\caption{(a) $U_c$ as function of $\mu$ within Stoner calculation with
  approximate band (solid red line), full band (dashed blue line) and
  the result of including vertex corrections (black triangles) for
  $\gamma_0=\Delta$ and
(a) $\gamma_1=2\Delta$ and (b) $\gamma_1=\Delta$. }
\label{fig:3}
\end{figure}

The critical coupling $U_c$ as
a function of $\mu$ in this vertex corrected scenario is shown in
Figs.~\ref{fig:3} (a) and (b)  (black
triangles). We find that the critical coupling
increases from the simple Stoner prediction, but the vertex corrected
values can be easily accessed experimentally in cold atom systems. We
also see that the vertex corrected critical coupling still decreases to zero as
the density is decreased towards half-filling and the instability is robust with respect
to vertex renormalizations.

  We can also
estimate the time-scale for formation of the ferromagnetic domains by
considering the location of the pole of the susceptibility on the
positive imaginary axis of the complex frequency plane when
interactions are tuned beyond the instability. We find this time-scale
for the growth of the domains to be $\sim 10^{-5} s$, which is much shorter
than the atom loss time scale $\sim 1 s$~\cite{doublondecay} in these systems. Thus
this instability should be clearly visible in the experiments.

Thus the Stoner instability is robust to the vertex corrections. In the
next section, we will see that the ladder diagrams represent the most
important corrections to the interaction vertex in a perturbative
renormalization group procedure, and thus we have shown that the
Stoner instability is robust to such corrections.

\subsection{Stability of the Incommensurate SDW state}

It is well known~\cite{BelitzRMP,Vojta1,Simmons,Conduit} that the presence of gapless fermionic particle-hole
excitations and the coupling of the order parameter to these soft
modes leads to a first order phase transition in the case of a Stoner
transition to a usual ferromagnetic state with a $q=0$ order
parameter. In certain parameter regimes, this first order transition
is pre-empted by a continuous transition to a spin-spiral state with a
broken translational symmetry~\cite{Vojta1}. In this section, we consider the soft
fermionic modes that our incommensurate SDW order parameter couples to
and show that they do not lead to a pre-emption of the Stoner
transition in this case of a finite $q=\delta k$ order parameter.

In case of the uniform ferromagnetic order, the order parameter
couples to particle-hole excitations at small wavevectors
$q\rightarrow 0$ and low frequencies $\omega \rightarrow 0$. The
density of states of these excitations have a singular behaviour in
this case $\sim \omega/q$, which invalidates the usual Landau Ginzburg
expansion around a continuous transition and leads to a first order
transition. In our case, the low energy theory is dominated by the
coupling of  the order parameter $\vec{m}_q$ to the soft fermionic
modes at $q \rightarrow\delta k$ and $\omega\rightarrow 0$ through a
term in the action
\beq
S \sim g \int d\tau m^+_q(\tau) \sum_k \psi^\dagger_{\downarrow k}(\tau)\psi_{\uparrow k+q}(\tau) +h.c.
\eeq   
where, for the sake of concreteness, we are considering the
order-parameter to lie in the $x-y$ plane, $\psi_{\sigma k}$ is the
fermion annihilation operator with spin $\sigma$ and momentum $k$ and
$g$ is a coupling parameter. The density of these excitations can be
obtained from the imaginary part of the momentum and frequency
dependent retarded polarization
function $\Pi(q,\omega+i0^+)$ in this system. We use the one band model
with the quartic dispersion, $\epsilon_k =\epsilon_0(k^2-k_0^2)^2$, which was used in the previous section to
compute this function and get
\begin{widetext}
\beq
\Pi(q,\omega) = -\frac{1}{2\pi^2} \int_{k_-}^{k_+} k dk \int_0^\pi
d\phi \left[\frac{1}{\omega + i0^+ -\epsilon_0\alpha(2k^2+\alpha-k_0^2)} -\frac{1}{\omega + i0^+
    -\epsilon_0\beta(2k^2-\beta-k_0^2)}\right]
\eeq 
where $\alpha=2kq \cos \phi +q^2$ and $\beta=2kq\cos \phi -q^2$.
\end{widetext}

We have explicitly calculated the full frequency and momentum
dependent polarization function near $q=\delta k$ and $\omega=0$ and
find (both numerically and analytically)  that the real part $\Pi^{'} \sim const + \sqrt{|q-\delta k|} $
whereas $\Pi^{''} \sim \omega$. We note that unlike the case of usual uniform
($q=0$) order parameter, the density of excitations does not have any
non-analyticity. Furthermore the finite $q$ order parameter couples in an
off-diagonal channel and hence the incommensurate density wave is not
unstable to the soft fermionic modes. 
We note that we are considering
a region in phase space, where there is no instability towards uniform
magnetism and the fluctuations of the uniform order parameter is
gapped on a sufficiently large scale (in fact this scale can be made
as large as possible by going closer to the Van-Hove singularity). Thus, unlike the uniform
magnetic order, the incommensurate SDW does not lead to a first order
transition. Furthermore, since there is no instability toward uniform
order, there is no chance of obtaining a spin-spiral where a
transverse propagating spin mode is superimposed on a uniform
background. The incommensurate SDW, of course leads to spatial
oscillation of the order-parameter in the system. Finally, we have not
seen any evidence of formation of nematic ordered states, although
such quadrupolar ordering cannot be ruled out completely. In the next
section, we construct an RG flow, which does not show any tendency
towards Pomeranchuk instabilities, although a truly spontaneously
broken nematic order is beyond the purview of such calculations.

\section{\label{sec:RG}Renormalization Group and Competing Orders}

In the previous sections, we have shown that the system with two Fermi
surfaces possess a Stoner-like instability towards an incommensurate
spin-density-wave order and this instability is robust to
incorporating particle-particle correlations through vertex
corrections. Since Stoner instability is a threshold phenomenon,
i.e. it requires a finite coupling before the instability occurs, it
is legitimate to ask whether this is pre-empted by an instability at
infinitesimal coupling towards a competing order like charge density wave or
superconductivity. Such perturbative instabilities, if they are
present, would drive the system towards these alternate symmetry
broken states. The formalism which treats competing orders on an
equal footing is renormalization group analysis, and it can pick out relevant orders and the relative strength of
instabilities in an unbiased way. 

In the case of fermions with one Fermi surface, it is
known~\cite{ShankarRG} that in dimensions more than one, the only
relevant instability is the superconducting instability with
infinitesimal attractive interactions, unless the Fermi surface is
nested~\cite{Honerkamp,MetznerRMP}, when density wave instabilities can occur. The
case of fermions with two Fermi surfaces~\cite{Nk1,Nk2}, well separated in momentum space has been worked out in the
context of pnictide
superconductors~\cite{pnictide,Chubukov1,Chubukov2}, where a
competition between a spin-density-wave order and superconductivity is
seen. In that case, the presence of a hole pocket and an electron
pocket, with Fermi velocities of opposite signs, lead to one loop
renormalization of both the interaction vertex which drives SDW and
the vertex which drives superconductivity and depending on the details
of the system, different orders can prevail in the system.

\begin{figure}
\includegraphics[width =0.45\linewidth]{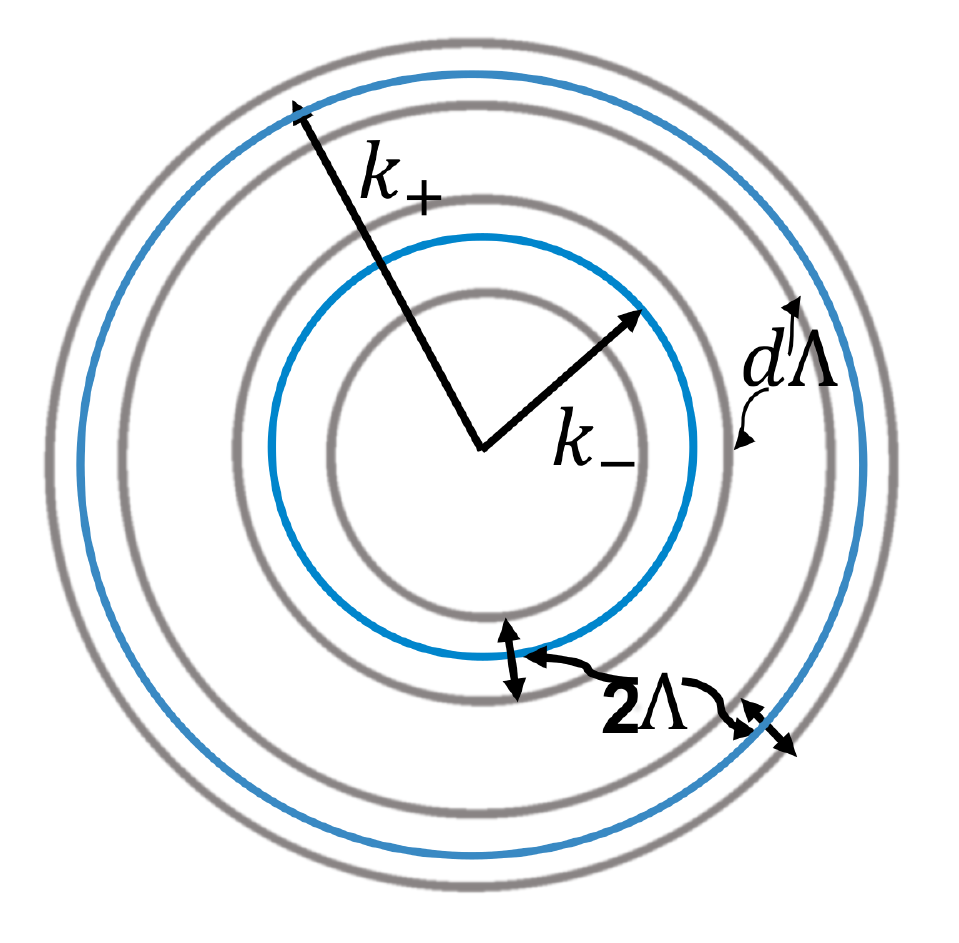}
\includegraphics[width =0.45\linewidth]{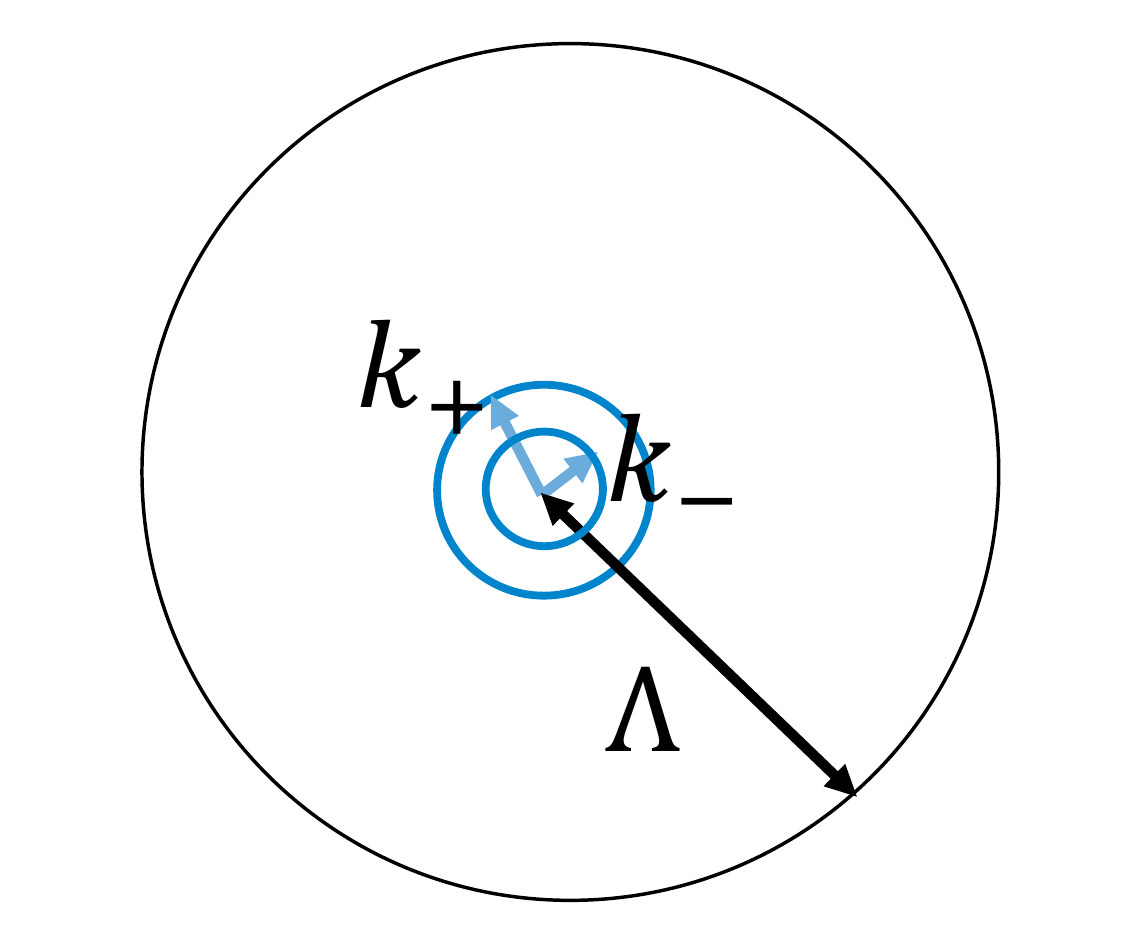}
\includegraphics[width =0.8\linewidth]{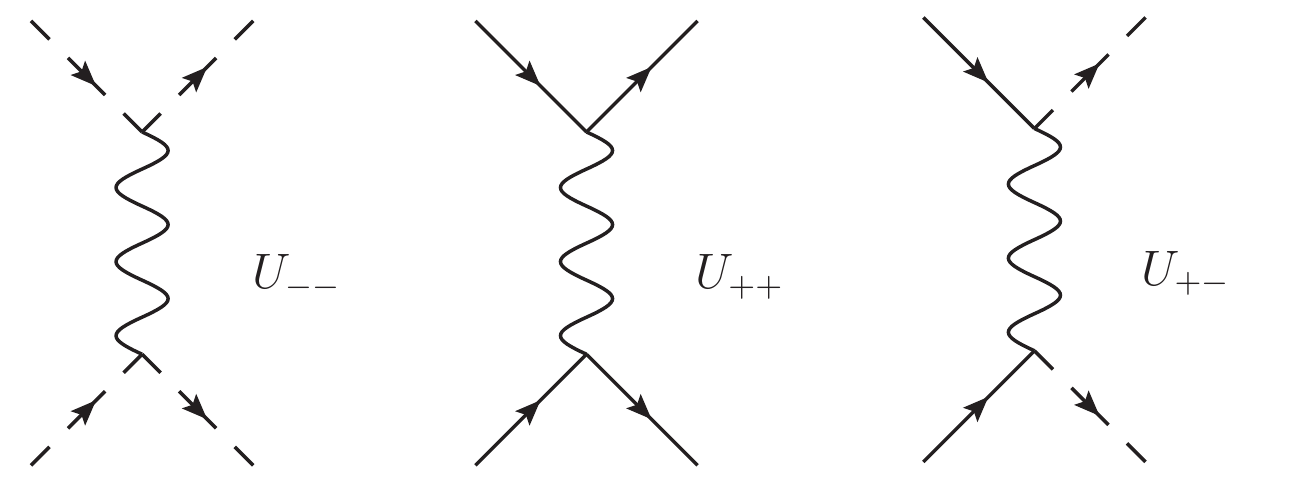}
\caption{Schematic sketch of cutoffs and momentum shells being
  integrated out in RG when (a) $\Lambda < \delta k/2$ and (b)
  $\Lambda \gg k_{\pm}$. (c) The Feynman vertices for the bare couplings
  showing intra and inter Fermi surface scatterings.}
\label{fig:RG}
\end{figure}
We would first like to point out that the geometry of our two fermi
surfaces is quite different from that found in pnictides, which has two small Fermi surfaces
separated by a large momentum in the Brillouin zone. In our case, we
have an annular Fermi sea with two concentric circles as the Fermi
surfaces. Let us first consider a cutoff scale $\Lambda \gg k_0,
\Lambda \gg k_{\pm}$. In this case, it is impossible to unambiguously
define electrons around the two Fermi surfaces, and for all practical
purposes, the system behaves as if it had one Fermi surface. This
situation is shown in Fig.~\ref{fig:RG} (b). In this case, the
standard RG of Fermions with a single Fermi surface~\cite{ShankarRG}
holds and both the forward scattering amplitude and the interaction
in the particle-particle channel are marginal at tree level and, for
repulsive interactions, renormalizes down logarithmically. This
logarithmic reduction is precisely what is captured by the ladder type
vertex corrections shown in the previous section. This does not lead
to any perturbative instabilities by itself, while the Stoner
instability has been shown to be robust to these corrections. The key
reason for this is that while the reduction due to RG is logarithmic,
the threshold required to attain the Stoner instability decreases as the square root of the energy scale (Fermi energy) and hence prevails over
the logarithmic reduction.

\begin{figure}
\includegraphics[width =0.8\linewidth]{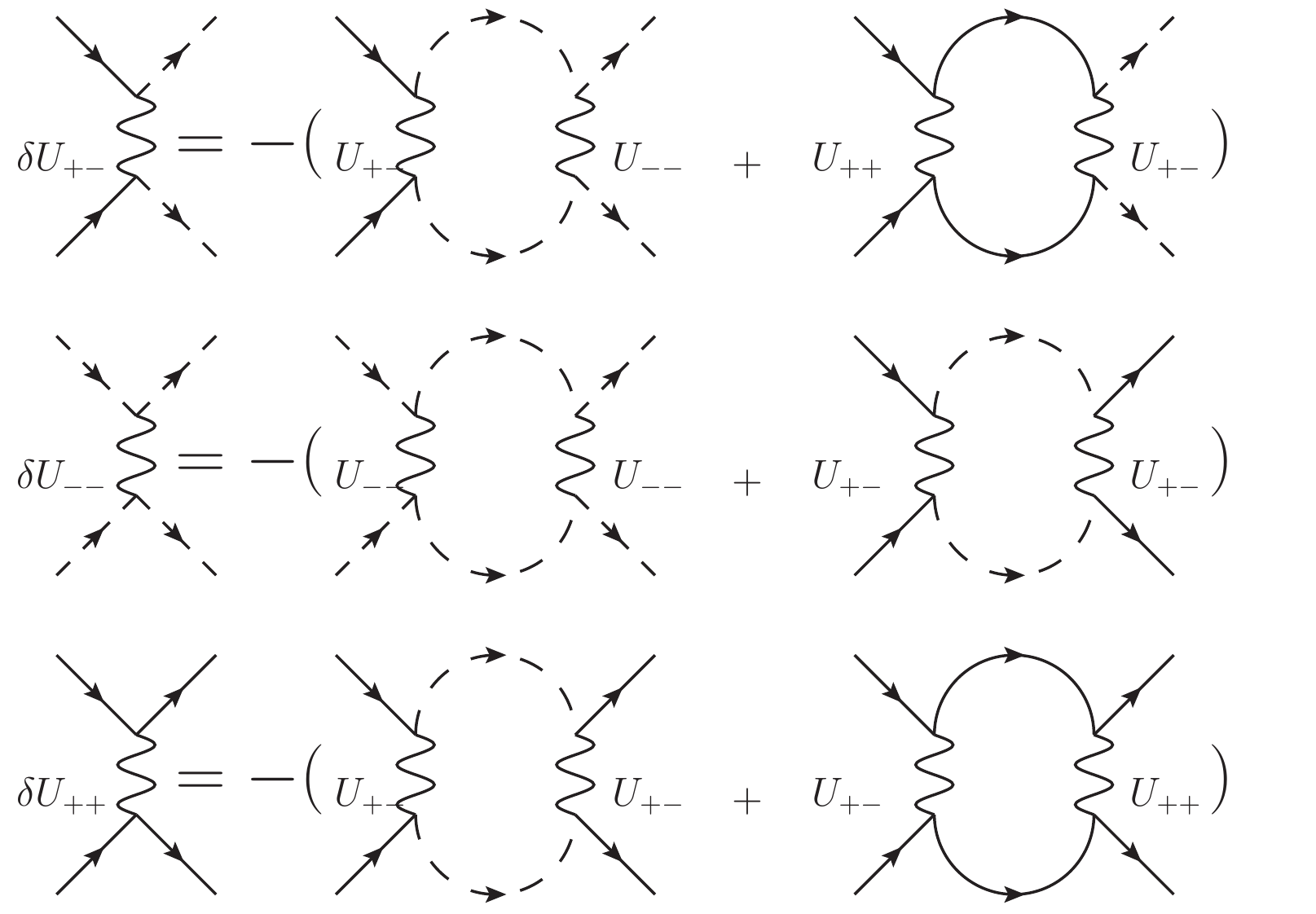}
\includegraphics[width =0.45\linewidth]{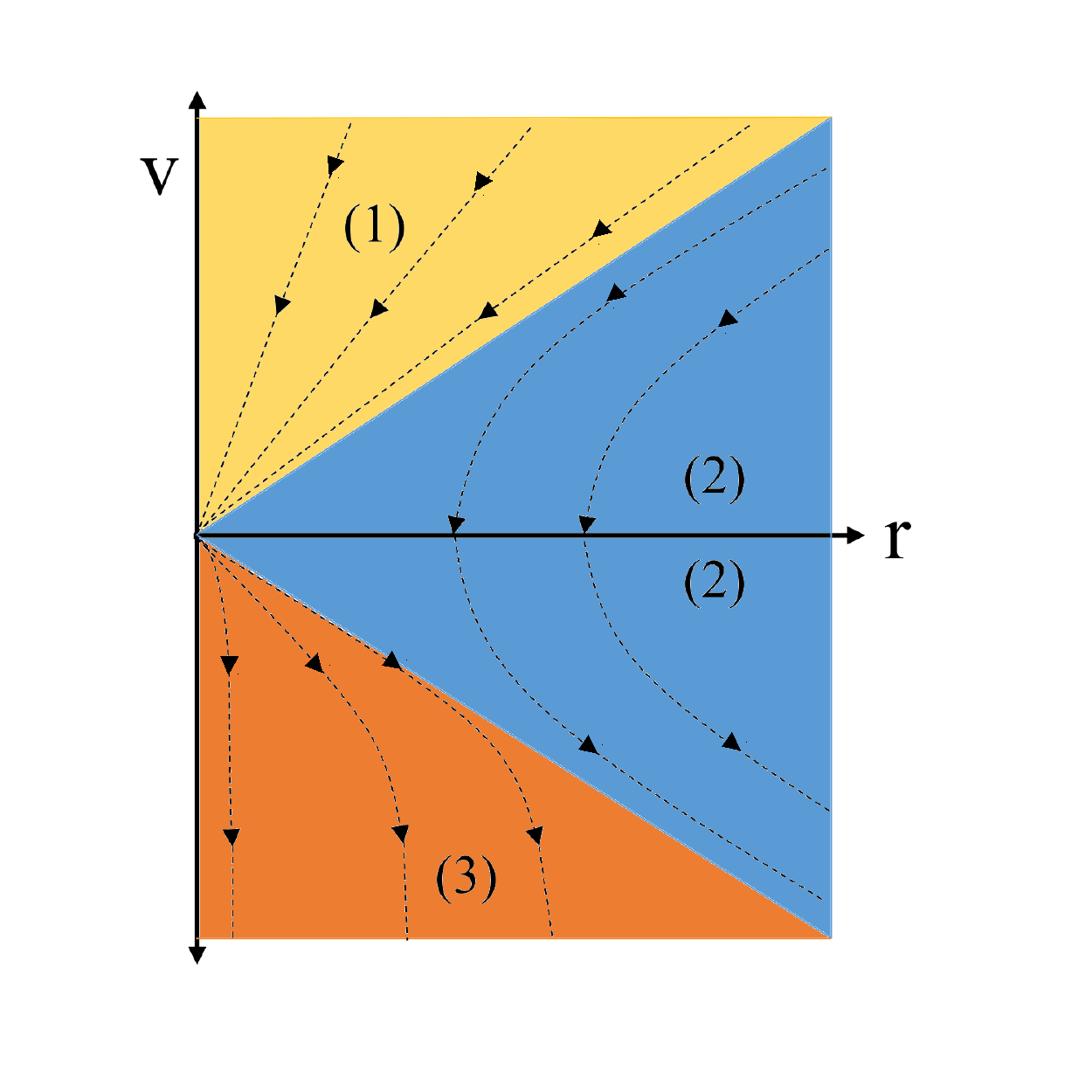}
\includegraphics[width =0.4\linewidth]{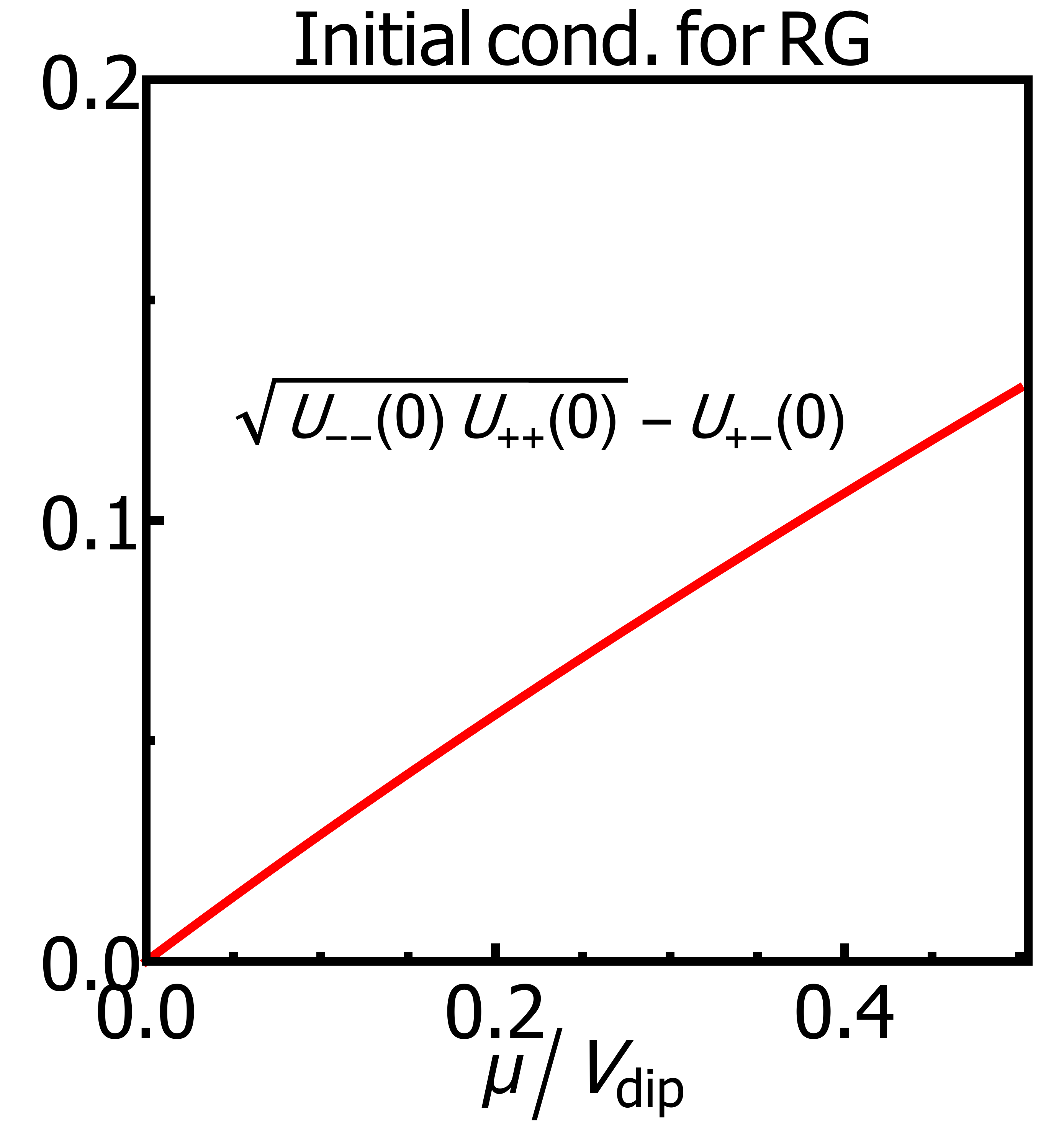}
\caption{(a) Feynman diagrams corresponding to one loop correction of
  the interaction vertices. (b) The RG flow diagram showing regions
  where the Fermi liquid is stable and regions where a superconducting
  instability takes over. (c) Calculated initial conditions showing
  that the system lies in the stable region and hence does not show a
  perturbative superconducting instability. Thus the Stoner threshold
  behaviour is not destroyed by any competing perturbative instability.}
\label{fig:RG2}
\end{figure}

A clear distinction between the two Fermi surfaces can be made when
$\Lambda < \delta k/2$, when one can unambiguously talk about low
energy fermions around the two Fermi surfaces, as shown in
Fig.~\ref{fig:RG}(a). In this case, the renormalization group
calculation proceeds along the lines of Ref.~\cite{Chubukov1}. The
key interaction vertices in the different channels are shown in
Fig.~\ref{fig:RG}(c). Here $U_{++}$ represents a process where two
fermions with total momentum $0$ and lying near the Fermi surface $+$
are scattered two other states near $+$ with zero net
momentum. $U_{--}$ is the same process with all the fermions near the
$-$ Fermi surface and $U_{+-}$ is the process where a total zero
momentum pair is scattered from $+$ to $-$ Fermi surface. In addition
there are forward scattering processes which are not shown as they are
not renormalized in one loop RG. In principle the particle-hole
diagrams can also pick up a logarithmic divergence due to the change
in the sign of the Fermi velocity between the Fermi surfaces. However, a key
difference from  the RG flows for pnictides is that in this case,
since the Fermi surfaces are concentric, these couplings do not flow
under one loop RG due to phase space restrictions, coming from
constraints of momentum conservation around the Fermi surfaces. 
The RG flow for the intra and inter Fermi surface BCS type dimensionless
couplings $U_{++}$, $U_{--}$ and $U_{+-}=U_{-+}$ are given by
\beq
\frac{d\hat{U}}{dl}=-\hat{U}^2
\eeq
 where $\hat{U}$ is a $2\times 2$ matrix $\left(\begin{array}{cc} U_{++} &
     U_{+-}\\U_{-+}& U_{--}\end{array}\right)$ and $ dl=d\Lambda/\Lambda $. The Feynman diagrams
 corresponding to this flow equation are shown in
 Fig.~\ref{fig:RG2}(a). We
 would like to note that the Fermi velocities at the 2 Fermi surfaces are
 not equal ($v_{+} \neq v_{-}$), but the factor of $v_{+}$ or
 $v_{-}$ drops out when the couplings are made dimensionless using
 the density of states at the respective Fermi surfaces, $ N_0=(1/2\pi)k_+/v_+=(1/2\pi)k_-/v_- $. These
 equations can be solved analytically to obtain the flow diagram of
 the couplings. The equations are most easily solved by variable
 transform to $\phi=2U_{+-}/(U_{++}-U_{--})$, $v=(U_{++}+U_{--})/2$
 and $r=\sqrt{U^2_{+-}+(U_{++}-U_{--})^2/4}$. It can be easily shown that $\phi$ does not flow
 under RG, while the flow patterns in the $v-r$ plane is shown in Fig.~\ref{fig:RG2}(b).
The half-plane is divided into three distinct regions by the lines
$v=r$ and $v=-r$. For $v>0$, if $v>r$ (yellow region (I)), the interaction parameters are
all renormalized to zero under RG and the Fermi liquid does not have
any perturbative instabilities.  For $v<0$, $v<r$ (orange region (III)), the flow moves away
from the Fermi liquid fixed point at the origin. This corresponds to
standard attractive superconducting instabilities, and since $|v|>r$,
the pairing will be dominated by the intra Fermi surface pairing. In
the blue region (II), $|v|<r$. If the flow starts from $v>0$, the
interaction scales are initially renormalized downwards, before
growing in the negative direction with $v=-r$ and $r \rightarrow
\infty$. This flow needs to be cutoff when $r\sim 1$ to obtain
critical points. We note that this case corresponds to the $s^\pm$
superconducting pairing instability, where the pairing symmetry is
$s-wave$ on both Fermi surfaces, but the pairing function changes sign
between the Fermi surfaces. Thus, we find that the only perturbative
instability in the system is towards superconductivity, provided the
flow starts from either the blue or the orange regions in the phase
diagram shown in Fig.~\ref{fig:RG} (b). We note that there is no instability
towards triplet superconductivity within this RG.

It is then important to determine where our system lies in this phase
diagram. The system will have a perturbative superconducting
instability provided $ |v|<r$, i.e. $U_{+-} <
\sqrt{|U_{++}U_{--}|}$. We first estimate the interactions at the
microscopic scale.  The microscopic values of the
  intra and inter Fermi surface couplings, obtained from the
  Hubbard U and the band wavefunctions, are shown in
  section ~\ref{sec:cold}. To calculate this, it is useful to work in
  the momentum space
\begin{equation}
H_{\text{int}} = U \sum_{\tau,k,k',q}c^{\dagger}_\tau(k)c_\tau(k+q)c^{\dagger}_\tau(k')c_\tau(k'-q)
\end{equation}. Here $ \tau $ is the layer/sublattice index and from
(\ref{eq:wavefn}) $
c^{\dagger}_\tau(k)=\phi^-_+(k)_{[\tau]}a^{\dagger}(k) $, where $a_k$
is the low energy electron operator in the band basis (we consider
only the band where the chemical potential lies). Projecting onto the
band basis and working with momenta on the respective Fermi surfaces,
one can easily construct $U_{++}$, $U_{--}$ and $U_{+-}$ for the system. We find that the system is always in
the regime where the Fermi liquid is stable to perturbative
instabilities. In Fig~\ref{fig:RG2}(c), we plot  $
\sqrt{|U_{++}U_{--}|}-U_{+-} $ as a function of the chemical potential and show
that this is always positive. This remains true under the weak
logarithmic reduction from the RG with large cutoff (which is similar
to RG with one Fermi surface), and thus the Stoner instability is not
overtaken by any other perturbative instabilities. We note that the
Stoner instability remains a threshold phenomenon, i.e. a finite
coupling strength is required to reach this instability, and is not
captured by the perturbative RG that we have constructed. We would
also like to note that there is an intermediate scale from $\Lambda
\sim k_\pm$ to $\Lambda \sim \delta k/2$, where the RG procedure cannot
be carried out since the non-linearities of the dispersions do not
allow a reasonable scaling analysis to start the procedure. 

We would also like to note that if we start from the insulating phase,
i.e. at the band-bottom, the effective dispersion $\sim k^2$, and this
would lead to non-Fermi liquid behaviour. However, chemical potential
is a relevant potential at this non-Fermi liquid fixed point and the
system flows to a Fermi liquid at any finite chemical potential, which
then undergoes the Stoner instability described in the previous
sections. This also clearly shows that the finite chemical potential
should be described in terms of an itinerant system and its
instabilities rather than as an insulating system.

Finally, having dispelled off the possibility of perturbative
instabilities, we have also checked that competing orders like charge
density wave, spin spirals and superconductivity of different orbital
symmetry do not win over the
incommensurate spin-density wave instability in the system. We have
also checked for the possibility of Pomeranchuk type instabilities and
have not found any of the Pomeranchuk channels to be more unstable
than the SDW instability. We note that we have only checked for order
parameters which are bilinear in the fermion fields, which leaves open
the possibility of a nematic order, which is impossible to pick up in
a one loop perturbative RG calculation.  However, the nematic
  order, which involves breaking down the $C_6$ symmetry of the lattice
  to a $C_2$ symmetry, would require the inter-valley and
  intra-valley effective interactions to be comparable. We have
  earlier shown, both within RPA and with inclusion of vertex
  corrections, that the inter-valley effective interactions are better
  screened and hence much
  weaker than intra-valley effective interactions. Thus a nematic
  instability is unlikely to occur in the case where the finite
  density of states leads to differential screening of the
  intra-valley and inter-valley interactions. In fact, if lattice effects (trigonal distortions) are included,
  the main effect of the imposition of discrete lattice symmetry is
  that the SDW wave-vector will lose the rotational symmetry and would
  choose between three degenerate vectors dictated by the effective $C_3$
  symmetry around each Dirac valley.

\section{\label{sec:conclusion}Conclusion and Discussions} 
Itinerant fermions with multiple Fermi surfaces can lead to
interesting phenomenology in a many body system. The biased Bernal
stacked bilayer honeycomb lattice provides such a system at densities
close to half-filling, with an annular Fermi sea.
In this paper, we have proposed a simple scheme of
implementing a Bernal-stacked bilayer lattice with ultracold atoms,
which is a modification of schemes already used by experimentalists to
implement a planar honeycomb lattice. The low energy band dispersion
in this system has the shape of a sombrero and results in an annular
Fermi sea with concentric Fermi surfaces at low densities. We have
used band structure calculations to correlate the optical lattice
parameters to the tight binding parameters of the problem. We have
shown that there is a wide range of experimentally accessible
parameters, where the effects of the annular Fermi sea can be
clearly seen.

The sombrero like dispersion at low densities, and the consequent
presence of two
Fermi wave-vectors in the system leads to singularities
in the static polarizability function of the system, which controls
the response of the system to potential perturbations. We find that
naive phase space arguments would predict $4$ singularities, whereas
$3$ are actually seen in the calculations. The absence of the fourth
singularity at a wave-vector equal to the sum of the two Fermi wave
vectors can be explained through a subtle cancellation of contribution
from two Fermi surfaces. The singularities in the static
polarizability is reflected in the occurrence of Friedel and RKKY
oscillations with three wave-vectors, which can be seen
experimentally, thus providing an evidence of the presence of multiple
Fermi wave-vectors.

If the fermions placed in this lattice are interacting  with a Hubbard repulsion, there is a quantum phase transition to an incommensurate
spin density wave state with a wave-vector equal to the thickness of
the Fermi sea. Within a simple Stoner approximation, the critical coupling for this transition goes to zero
as the density of the system is lowered and the chemical potential
approaches the Van Hove singularity at the band bottom. The inclusion of
vertex corrections, which are the leading order correction from a
renormalization group analysis at large cutoff, does not change this
picture substantially, with small quantitative changes in $ U_c $. The
Stoner instability is thus robust to many body renormalization of the
couplings between fermions. We note that by virtue of the azimuthal symmetry of our low energy
model, the modulus of the incommensurate wave-vector is specified
while its direction is indeterminate within our calculation. In
reality, we have neglected small lattice effects like trigonal
warping,  which will break the circular symmetry, and choose a
direction with a three-fold degeneracy respecting the lattice symmetry
of the full model.

We also construct a perturbative renormalization group flow for this
system to treat competing order parameters on an equivalent
footing. We show that the large cutoff RG leads to a
logarithmic reduction of the interaction strength, well understood in
the RG of fermions with a single Fermi surface. This is in contrast
with RG flow for two Fermi surface systems like iron pnictides, where
the RG flow leads to increasing couplings and a non-trivial fixed
point. The main reason our RG flow follows that of a single Fermi
surface is that unlike pnictides, which have small Fermi seas separated
by a large almost commensurate wave-vector, our Fermi surfaces are
concentric, and at large cutoff, the system behaves as if there is
only one Fermi surface. At small cutoffs, where the two Fermi surfaces
can be clearly distinguished, there is a superconducting instability
in parts of the parameter regime, but we show that our system falls
outside this parameter regime, and hence there are no perturbative
instabilities. Thus in absence of perturbative instabilities, the
Stoner instability towards SDW states prevail over other
instabilities to remain the dominant threshold instability. We note
that we have considered competing orders which are bilinear in the
fermion fields, and a nematic instability remains a possibility, which
we cannot probe with our approach. We would also like to note that while our RG procedure is
well defined for very large and very small cutoffs, there is a range
of scales from $k_{+}$ to $\delta k$, where it is not possible to do
a reliable RG calculation. Within these caveats, we find the Stoner
instability towards the incommensurate spin-density-wave state to be
the dominant instability of the system.
 
The predicted incommensurate spin
density wave order can be identified by
different experimental techniques. Noise measurements~\cite{noise} and spectroscopic techniques like spin-dependent
Bragg spectroscopy~\cite{bragg} and optical lattice modulation spectroscopy~\cite{optlat} can be used to look at the finite wave vector spin
order. Alternatively, a quantum quench across the critical coupling
would lead to the growth of magnetic domains~\cite{Pekker} with a length scale which
corresponds to the most unstable mode. In this case, the growth of the most unstable
mode at $q=\delta k$ can be observed using real space imaging techniques.

\section{Acknowledgments}
R. S. wishes to thank Sankar Das Sarma, David Pekker and Kazi Rajibul
Islam for useful discussions. S. D. wishes to thank Tomi H. Johnson for
help with the MLGWS software package. R.S. and S.D acknowledge computational facilities
at the Dept. of Theoretical Physics, TIFR Mumbai.



\begin{thebibliography}{99}
\bibitem{footnote1} In addition, there is a weak tunneling $\gamma_3$
  between $A_1-A_2$ and $B_1-B_2$ , and $\gamma_4$ between
  $A_2-B_1$. In this paper we have set these weak tunneling parameters
  to $0$.
\bibitem{footnote2} There is a factor of 2 difference in the
  definition of the lattice potential between our convention and that
  of Ref.~\cite{coldatom_bandparam_2}.
\bibitem{footnote3} We note that the spin degeneracy is not involved
  in the calculation of the spin susceptibility.
\end{thebibliography}
\end{document}